\documentclass[journal]{IEEEtran}
\usepackage{amsmath,amsfonts,amssymb}
\usepackage{graphics}
\usepackage{epsfig}
\usepackage{xcolor}
\usepackage{url}
\usepackage{multirow}

\usepackage{amsthm}
\theoremstyle{definition}
\newtheorem{lemma}{\normalfont \bfseries Lemma}

\newtheorem{theorem}{\normalfont\bfseries Theorem}
\newtheorem{assumption}{\normalfont\bfseries Assumption}

\newtheorem{definition}{\normalfont\bfseries Definition}

\newtheorem*{problem}{\normalfont \bfseries Problem Statement}

\newcommand{\mb}[1]{\mathbf{#1}}
\newcommand{\bs}[1]{\boldsymbol{#1}}

\newcommand{\xx}{\mb{x}}
\newcommand{\uu}{\mb{u}}
\newcommand{\ff}{\mb{f}}
\renewcommand{\gg}{\mb{g}}
\newcommand{\kk}{\mb{k}}
\newcommand{\kd}{\kk_{\rm d}}
\newcommand{\GGamma}{\bs{\Gamma}}
\newcommand{\WW}{\mb{W}}
\newcommand{\bb}{\mb{b}}

\newcommand{\R}{\mathbb{R}}
\newcommand{\X}{\mathbb{R}^n}
\newcommand{\U}{\mathbb{R}^m}
\newcommand{\C}{\mathcal{C}}
\newcommand{\K}{\mathcal{K}}

\newcommand{\Kinf}{\K_\infty}
\newcommand{\Keinf}{\K_\infty^{\rm e}}
\newcommand{\derp}[2]{\frac{\partial #1}{\partial #2}}
\newcommand{\II}{\mb{I}}
\newcommand{\OO}{\mb{0}}

\renewcommand{\c}{{\rm c}}
\newcommand{\s}{{\rm s}}
\renewcommand{\t}{{\rm t}}

\newcommand{\rr}{\mb{r}}
\newcommand{\xxi}{\bs{\xi}}
\newcommand{\eeta}{\bs{\eta}}
\newcommand{\oomega}{\bs{\omega}}
\newcommand{\vv}{\mb{v}}
\renewcommand{\aa}{\mb{a}}
\newcommand{\VT}{V_{\rm T}}
\newcommand{\AT}{A_{\rm T}}
\newcommand{\zzeta}{\bs{\zeta}}

\newcommand{\Reb}{\mb{R}_{\rm eb}}
\newcommand{\Rbe}{\mb{R}_{\rm be}}
\newcommand{\Rbw}{\mb{R}_{\rm bw}}
\newcommand{\Rwb}{\mb{R}_{\rm wb}}
\newcommand{\vw}{\vv_{\rm w}}
\newcommand{\vb}{\vv_{\rm b}}
\newcommand{\HH}{\mb{H}}
\newcommand{\FT}{F_{\rm T}}
\newcommand{\MM}{\mb{M}}
\newcommand{\FFT}{\mb{F}_{\rm T}}
\newcommand{\FA}{\mb{F}_{\rm A}}
\newcommand{\ggD}{\gg_{\rm D}}
\newcommand{\gD}{g_{\rm D}}
\renewcommand{\AA}{\mb{A}}
\newcommand{\JJ}{\mb{J}}

\newcommand{\vvd}{\vv_{\rm d}}
\newcommand{\vs}{\vv_{\rm s}}
\newcommand{\ad}{\aa_{\rm d}}
\newcommand{\as}{\aa_{\rm s}}
\newcommand{\Rd}{R_{\rm d}}
\newcommand{\Rs}{R_{\rm s}}
\newcommand{\KK}{\mb{K}}
\newcommand{\vc}{\vv_{\rm c}}
\newcommand{\ac}{\aa_{\rm c}}

\renewcommand{\S}{\mathcal{S}}
\newcommand{\hp}{h_{\rm p}}
\newcommand{\hpi}{h_{{\rm p},i}}
\newcommand{\he}{h_{\rm e}}
\newcommand{\hei}{h_{{\rm e},i}}
\newcommand{\hb}{h_{\rm b}}

\newcommand{\nn}{\mb{n}}
\newcommand{\PP}{\mb{P}}
\newcommand{\hV}{h_{V}}

\begin{document}

\title{Collision Avoidance and Geofencing for Fixed-wing Aircraft with Control Barrier Functions}

\author{Tamas G. Molnar,~\IEEEmembership{Member,~IEEE},
Suresh K. Kannan,
James Cunningham,
Kyle Dunlap,
Kerianne L. Hobbs, and
Aaron D. Ames,~\IEEEmembership{Fellow,~IEEE}
\thanks{*Approved for public release: distribution is unlimited. Case Number AFRL-2024-0947, AFRL-2024-2450. The views expressed are those of the authors and do not reflect the official guidance or position of the United States Government, the Department of Defense or of the United States Air Force. This research is supported in part by the National Science Foundation (CPS Award \#1932091), AFOSR, and Nodein Inc. through USAF Award \#FA864922P0787.}%
\thanks{T. G. Molnar is with the Department of Mechanical Engineering, Wichita State University, Wichita, KS 67260, USA, {\tt\small tamas.molnar@wichita.edu}.}%
\thanks{S. K. Kannan is with the Nodein Autonomy Corporation, Farmington, CT 06085, USA, {\tt\small kannan@nodein.com}.}%
\thanks{J. Cunningham is with Parallax Advanced Research, Beavercreek, OH 45431, USA, {\tt\small james.cunningham@parallaxresearch.org}.}%
\thanks{K. Dunlap is with Parallax Advanced Research, Beavercreek, OH 45431, USA, {\tt\small kyle.dunlap@parallaxresearch.org}.}%
\thanks{K. L. Hobbs is with the Autonomy Capability Team (ACT3), Air Force Research Laboratory, Wright-Patterson AFB, OH 45433, USA, {\tt\small kerianne.hobbs@us.af.mil}.}%
\thanks{A. D. Ames is with the Department of Mechanical and Civil Engineering, California Institute of Technology, Pasadena, CA 91125, USA, {\tt\small ames@caltech.edu}.}%
}

\markboth{IEEE Transactions on Control Systems Technology}%
{Molnar et al.: Collision Avoidance and Geofencing for Fixed-wing Aircraft with CBFs}

\maketitle


\begin{abstract}
Safety-critical failures often have fatal consequences in aerospace control.
Control systems on aircraft, therefore, must ensure the strict satisfaction of safety constraints, preferably with formal guarantees of safe behavior.
This paper establishes the safety-critical control of fixed-wing aircraft in collision avoidance and geofencing tasks.
A control framework is developed wherein a run-time assurance (RTA) system modulates the nominal flight controller of the aircraft whenever necessary to prevent it from colliding with other aircraft or crossing a boundary (geofence) in space.
The RTA is formulated as a safety filter using control barrier functions (CBFs) with formal guarantees of safe behavior.
CBFs are constructed and compared for a nonlinear kinematic fixed-wing aircraft model.
The proposed CBF-based controllers showcase the capability of safely executing simultaneous collision avoidance and geofencing, as demonstrated by simulations on the kinematic model and a high-fidelity dynamical model.
\end{abstract}

\begin{IEEEkeywords}
Aerospace control, Aircraft navigation, Control barrier function, Collision avoidance, Geofencing
\end{IEEEkeywords}

\section{Introduction}
\IEEEPARstart{S}{afe} behavior is of utmost importance for aerial vehicles due to the severity of consequences in case of an incidental failure.
Thus, on-board control systems must satisfy strict safety constraints while operating aircraft during flight.
Safety constraints may span a wide range of criteria, including bounds on the aircraft's flight envelope and control surfaces (envelope protection), altitude, speed, acceleration, load factors, orientation, angular rates, angle of attack, control surface deflections and their rates, as well as bounds on an aircraft's trajectory (safe navigation) that may include geofences (keep-in or keep-out zones) and collision avoidance with the ground or other air vehicles.
Such strict safety specifications necessitate control systems that are designed in a safety-critical fashion, preferably with guarantees or certificates of safe behavior under certain operating conditions.
Recently, this has led to the idea of run-time assurance (RTA) systems~\cite{hobbs2023rta} that serve as an added module to the aircraft's nominal flight controller and intervene whenever necessary to avoid the violation of safety.
RTA is often used to supervise complex control systems whose safe behavior is difficult to guarantee or verify, such as learning-based controllers~\cite{fuller2020rta, Nagarajan2021}.
RTA systems showed promising results in 
indoor flight with quadrotors~\cite{chaki2017certifiable},
navigation of unmanned fixed-wing aircraft along safe corridors~\cite{schierman2020rta},
neural network-based aircraft taxiing application~\cite{cofer2020rta}, and
aerial refueling task in naval aviation~\cite{costello2023rta}.
How to design safe control laws for use in RTA, however, is still an open problem.

\begin{figure}[!t]
\centering
\includegraphics[scale=1]{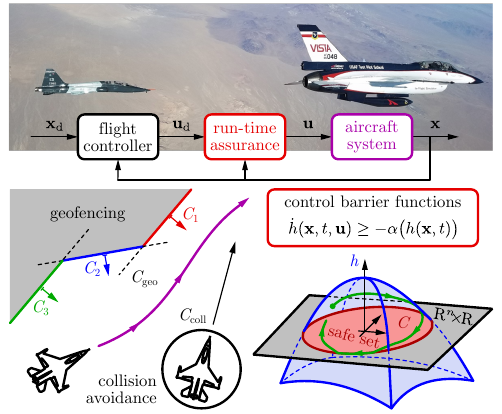}
\caption{Overview of the safety-critical control framework for fixed-wing aircraft.
The motion of a single aircraft is controlled to avoid collisions with other aircraft and prevent crossing a geofence in 3D space.
To this end, the proposed control barrier function-based run time assurance system intervenes into the nominal flight controller whenever necessary to avoid dangers.}
\label{fig:concept}
\end{figure}

Safety has long been of interest in the aerospace control literature.
Many studies have focused on safe aircraft navigation, where the overall motion is controlled in scenarios like collision avoidance, trajectory tracking, and geofencing.
Approaches include the use of potential fields~\cite{Khatib1985, Chen2016},
velocity obstacles~\cite{Fiorini1998}, control barrier functions (CBFs)~\cite{AmesXuGriTab2017}, controlled invariant sets~\cite{shoukry2017closedform}, and
reinforcement learning~\cite{Ravaioli2022, Hobbs2023}.
Notably, CBFs provide provable safety guarantees as opposed to most learning-based methods.
Furthermore, some CBF formulations can be used to generate control invariant sets~\cite{gurriet2020scalable}, while CBFs can even be constructed from artificial potential fields \cite{Singletary2021} and velocity obstacles~\cite{Haraldsen2023}.
Most CBF-based controllers respond to the current state of the system without a preview of future trajectories.
This may lead to suboptimal behavior compared to other methods with preview, such as motion planning algorithms or model predictive control, but it also makes CBF-based controllers easy to compute.

CBFs demonstrated success in collision avoidance on quadrotors in simulation~\cite{Tayal2023, Llanes2023} and experiments~\cite{Singletary2021}, as well as on fixed-wing aircraft in the context of probabilistic safety certificates~\cite{Luo2019}, learning-based~\cite{Scukins2021} and data-driven CBFs~\cite{squires2021modelfree}, and multi-aircraft control~\cite{squires2022composition}.
Trajectory tracking by fixed-wing aircraft also leveraged CBF theory to keep path following errors within prescribed bounds, via high-order CBFs~\cite{Zhou2020}, barrier Lyapunov functions~\cite{xu2022barrierlyapunov}, and robust CBFs~\cite{zheng2023constrained}.
Moreover, geofencing---where the aircraft is navigated to avoid a restricted territory beyond a ``geofence'' like the airspace of a country or private property---was also addressed by CBFs with success on quadrotors~\cite{Ghaffari2021, singletary2022onboard}.
A CBF-based fixed-wing aircraft RTA system for simultaneous collision avoidance and geofencing 
is yet to be developed.

Notably,~\cite{Corraro2022} addressed simultaneous collision avoidance and geofencing without CBFs, for tactical unmanned aircraft considering multiple intruder aircraft, fixed obstacles, no-fly zones, and bad weather areas.
An optimal control problem was formulated with risk assessment and probabilistic safety constraints.
In our present work, we instead focus on deterministic safety constraints, and develop a CBF-based RTA system for simultaneous collision avoidance and geofencing.
We use a reduced-order model of the aircraft dynamics to formulate the CBF and synthesize a controller in closed form that provides formal guarantees of safety.
The approach of developing safety-critical controllers via reduced-order models and CBFs has been successful on a variety of autonomous systems, including legged, wheeled, and flying robots, manipulators, and heavy-duty trucks~\cite{cohen2024reduced}.
Now we seek to extend this approach to establish RTA for fixed-wing aircraft.

In this paper, we propose a safety-critical RTA system for fixed-wing aircraft using CBFs, as summarized by Fig.~\ref{fig:concept}.
Specifically, we accomplish simultaneous collision avoidance and geofencing in 3D space on a nonlinear kinematic model of an aircraft.
To the best of our knowledge, this is the first application of CBFs to simultaneous collision avoidance and geofencing for fixed-wing aircraft.
First, we encode multiple safety constraints (related to collision avoidance and multiple geofence boundaries) into a single CBF candidate.
Second, we establish how to construct CBFs for nonlinear aircraft dynamics.
We introduce and compare three approaches: high-order CBFs, backstepping-based CBFs, and model-free CBFs.
We highlight that a careful CBF construction is required to leverage all control inputs and make the aircraft both accelerate, pitch, and roll for safe collision avoidance and geofencing.
This challenge can be addressed, for example, via backstepping, and this work is the first application of CBF backstepping to fixed-wing aircraft.
Finally, we demonstrate the behavior of the proposed RTA by simulations of the kinematic model and also a high-fidelity dynamical model.

The paper is structured as follows.
Section~\ref{sec:modeling} describes the nonlinear model of fixed-wing aircraft kinematics.
Section~\ref{sec:CBF} provides background on CBFs.
Section~\ref{sec:RTA} introduces the main contributions by establishing and simulating the proposed CBF-based RTA system.
Section~\ref{sec:concl} closes with conclusions.

\section{Modeling the Aircraft's Motion}
\label{sec:modeling}

First, we introduce a kinematic model, called {\em 3D Dubins model}, that describes the motion of fixed-wing aircraft.
Kinematic models are popular in navigation and path planning~\cite{LugoCardenas2014, Owen2015}.
This model is derived in Appendix~\ref{appdx:model} from six-degrees-of-freedom dynamics using certain assumptions (cf.~Assumption~\ref{assum:model}).
The main assumption is that the angles of attack and sideslip are negligible.
We use this model for safety-critical controller synthesis, to design acceleration and angular velocity commands for the aircraft in a provably safe fashion.

\begin{figure}[!t]
\centering
\includegraphics[scale=1]{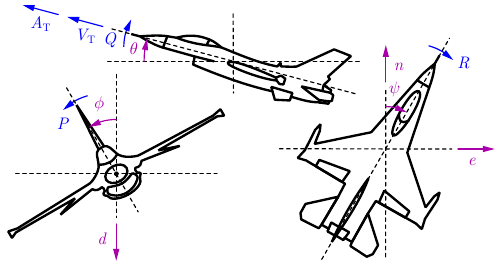}
\caption{Illustration of aircraft kinematics: position coordinates $n$, $e$, $d$, Euler angles $\phi$, $\theta$, $\psi$, speed $\VT$, angular velocities $P$, $Q$, $R$, and acceleration $\AT$.}
\label{fig:model}
\end{figure}

Consider the fixed-wing aircraft in Fig.~\ref{fig:model}.
We describe its motion by using
its: position along north, east, down directions, $n$, $e$, $d$;
roll, pitch, yaw Euler angles, $\phi$, $\theta$, $\psi$;
speed, $\VT$;
angular velocities about front, right, down axes, $P$, $Q$, $R$;
and longitudinal acceleration, $\AT$.
Specifically, we use the {\em 3D Dubins model} detailed in Appendix~\ref{appdx:model} as governing equation:
\begin{align}
\begin{split}
    \dot{n} & = \VT \c_{\psi} \c_{\theta}, \\
    \dot{e} & = \VT \s_{\psi} \c_{\theta}, \\
    \dot{d} & = -\VT \s_{\theta}, \\
    \dot{\phi} & = P + \s_{\phi} \t_{\theta} Q + \c_{\phi} \t_{\theta} R, \\
    \dot{\theta} & = \c_{\phi} Q - \s_{\phi} R, \\
    \dot{\psi} & = \frac{\s_{\phi}}{\c_{\theta}} Q + \frac{\c_{\phi}}{\c_{\theta}} R, \\
    \dot{V}_{\rm T} & = \AT,
\end{split}
\label{eq:Dubins}
\end{align}
where $\s_{(.)}$, $\c_{(.)}$, $\t_{(.)}$ stand for $\sin(.)$, $\cos(.)$, $\tan(.)$ and:
\begin{equation}
    R = \frac{\gD}{\VT} \s_{\phi} \c_{\theta},
\label{eq:Dubins_R}
\end{equation}
with $\gD$ being the gravitational acceleration.

The top 3 rows of~\eqref{eq:Dubins} give the velocity of the center of mass.
The next 3 rows of~\eqref{eq:Dubins} are the typical 3-2-1 Euler angle kinematics \cite{stevens2016} that relate orientation to angular velocity.
The last row of~\eqref{eq:Dubins} associates speed with acceleration.
Note that~\eqref{eq:Dubins} is true for the motion of any rigid body in an inertial frame.
Meanwhile,~\eqref{eq:Dubins_R} makes this an aircraft model by stating that the airplane must roll in order to turn left or right (${R \neq 0}$ requires ${\phi \neq 0}$).
The model has 7 states (3 positions, 3 orientation angles, and speed), while we consider 3 control inputs (acceleration and angular velocities related to rolling and pitching).
The state $\xx$ and input $\uu$ read:
\begin{align}
\begin{split}
    \xx & =
    \begin{bmatrix}
        n & e & d & \phi & \theta & \psi & \VT
    \end{bmatrix} ^\top, \\
    \uu & =
    \begin{bmatrix}
        \AT & P & Q
    \end{bmatrix}^\top.
\end{split}
\label{eq:stateinput}
\end{align}

To compress notations, we introduce the position vector $\rr$, the velocity-related states $\zzeta$, and the velocity vector $\vv$:
\begin{equation}
    \rr =
    \begin{bmatrix}
    n \\ e \\ d
    \end{bmatrix}, \quad
    \zzeta =
    \begin{bmatrix}
    \VT \\ \theta \\ \psi
    \end{bmatrix}, \quad
    \vv(\zzeta) =
    \begin{bmatrix}
        \VT \c_{\theta} \c_{\psi} \\
        \VT \c_{\theta} \s_{\psi} \\
        -\VT \s_{\theta}
    \end{bmatrix}.
\label{eq:posvel}
\end{equation}
With these, the dynamics~\eqref{eq:Dubins}-\eqref{eq:Dubins_R} become:
\begin{align}
\begin{split}
    \dot{\rr} & = \vv(\zzeta), \\
    \dot{\zzeta} & = \ff_{\zzeta}(\zzeta,\phi,R,\AT,Q), \quad
    R = \rho(\zzeta,\phi), \\
    \dot{\phi} & = f_{\phi}(\zzeta,\phi,R,Q,P).
\end{split}
\label{eq:Dubins_cascade}
\end{align}
These dynamics form a cascaded structure, where the evolution of higher-level states depends on lower-level states. 
Specifically, the evolution of the position $\rr$ is given by the velocity-related states $\zzeta$.
The evolution of $\zzeta$ depends on the remaining state $\phi$, the state-dependent turning rate $R$, and two control inputs $\AT$ and $Q$.
Finally, the evolution of the last state $\phi$ involves the third control input $P$.
Overall, the dynamics have a 3-level cascaded structure, where the control inputs enter at the 2nd and 3rd level (with an auxiliary algebraic formula for $R$).
Importantly, the right-hand side expressions:
\begin{align}
    \ff_{\zzeta}(\zzeta,\phi,R,\AT,Q) & =
    \begin{bmatrix}
        \AT \\
        \c_{\phi} Q - \s_{\phi} R \\
        \frac{\s_{\phi}}{\c_{\theta}} Q + \frac{\c_{\phi}}{\c_{\theta}} R
    \end{bmatrix}, \quad
    \rho(\zzeta,\phi) = \frac{\gD}{\VT} \s_{\phi} \c_{\theta}, \nonumber \\
    f_{\phi}(\zzeta,\phi,R,Q,P) & = P + \s_{\phi} \t_{\theta} Q + \c_{\phi} \t_{\theta} R
\end{align}
are affine in the turning rate $R$ and the control inputs $\AT$, $P$, $Q$.
We will leverage this structure when designing controllers.

Furthermore, we also write~\eqref{eq:Dubins}-\eqref{eq:Dubins_R} into the compact form:
\begin{equation}
    \dot{\xx} = \ff(\xx) + \gg(\xx) \uu,
\label{eq:Dubins_affine}
\end{equation}
where $\xx$ and $\uu$ are given in~\eqref{eq:stateinput} and:
\begin{align}
\begin{split}
    \ff(\xx) & =
    \begin{bmatrix}
    \vv(\zzeta) \\
    \ff_{\xxi}(\xx) \\
    0
    \end{bmatrix}, \quad
    \gg(\xx) =
    \begin{bmatrix}
    \multicolumn{3}{c}{\OO_{3 \times 3}} \\
    \multicolumn{3}{c}{\gg_{\xxi}(\xx)} \\
    1 & 0 & 0
    \end{bmatrix},
\end{split}
\end{align}
with:
\begin{align}
\begin{split}
    \ff_{\xxi}(\xx) = \frac{\gD}{\VT}
    \begin{bmatrix}
        \s_{\phi} \c_{\phi} \s_{\theta} \\
        - \s_{\phi}^2 \c_{\theta} \\
        \s_{\phi} \c_{\phi}
    \end{bmatrix}, \quad
    \gg_{\xxi}(\xx) =
    \begin{bmatrix}
        0 & 1 & \s_{\phi} \t_{\theta} \\
        0 & 0 & \c_{\phi} \\
        0 & 0 & \frac{\s_{\phi}}{\c_{\theta}}
    \end{bmatrix}.
\end{split}
\end{align}

In what follows, we synthesize controllers for the aircraft model~\eqref{eq:Dubins}-\eqref{eq:Dubins_R}, with the end goal of formal safety guarantees w.r.t.~collision avoidance and geofencing.
In particular, we design safe controllers for the control-affine form~\eqref{eq:Dubins_affine} of the model by using control barrier functions constructed based on the cascaded structure~\eqref{eq:Dubins_cascade} of the dynamics.
When designing the controller, we consider that the position $\rr$ of the aircraft is safe at time $t$ if the pair $(\rr,t)$ is inside a safe set ${\S \subset \R^3 \times \R_{\geq 0}}$ defined by the collision avoidance and geofencing constraints (see details in Section~\ref{sec:RTA}).

\begin{problem}
Consider system~\eqref{eq:Dubins_affine} and design a controller ${\kk : \X \times \R \to \U}$, ${\uu = \kk(\xx,t)}$ that guarantees safety w.r.t.~collision avoidance and geofencing.
Specifically, the controller must ensure that the trajectory $\rr(t)$ of the closed-loop system is maintained within a prescribed set $\S$: ${(\rr(t),t) \in \S}$, ${\forall t \geq 0}$ for initial positions satisfying ${(\rr_0,0) \in \S}$.
\end{problem}

\section{Control Barrier Functions}
\label{sec:CBF}

We utilize control barrier functions (CBFs) \cite{ames2014control} to guarantee safety on fixed-wing aircraft w.r.t.~collision avoidance and geofencing.
In particular, we consider time-dependent CBFs because safety depends on time when other aircraft move or the geofence is updated.

{\em Notations.}
If ${\alpha : \R \to \R}$ is continuous and strictly increasing with ${\alpha(0)=0}$ and ${\lim_{r \to \pm \infty} \alpha(r) = \pm \infty}$, then $\alpha$ is of extended class-$\Kinf$ (${\alpha \in \Keinf}$).
Norms are denoted by ${\|\uu\|^2 = \uu^\top \uu}$ and ${\|\uu\|_{\GGamma}^2 = \uu^\top \GGamma \uu}$ for ${\uu \in \R^m}$ and positive definite ${\GGamma \!\in\! \R^{m \times m}}$.
The zero vector in $\R^3$ is $\OO$, the zero matrix in $\R^{m \times n}$ is $\OO_{m \times n}$, and the identity matrix in $\R^{3 \times 3}$ is $\II$.

\subsection{Theoretical Background}
Motivated by the aircraft model~\eqref{eq:Dubins_affine}, consider control systems with state ${\xx \in \X}$ and control input ${\uu \in \U}$:
\begin{equation}
    \dot{\xx} = \ff(\xx) + \gg(\xx) \uu.
\label{eq:system}
\end{equation}
Let ${\ff : \X \to \X}$ and ${\gg : \X \to \R^{m \times n}}$ be locally Lipschitz continuous.
Given these dynamics, our goal is to design a locally Lipschitz continuous controller ${\kk : \X \times \R \to \U}$, ${\uu = \kk(\xx,t)}$ such that the closed control loop:
\begin{equation}
    \dot{\xx} = \ff(\xx) + \gg(\xx) \kk(\xx,t)
\label{eq:closedloop}
\end{equation}
exhibits safe behavior.
We characterize the behavior by the solution $\xx(t)$ of~\eqref{eq:closedloop} with initial condition ${\xx(0) = \xx_0 \in \X}$, and we assume that $\xx(t)$ exists for all ${t \geq 0}$.

We consider the system to be safe if the solution $\xx(t)$ at any time $t$ is kept within a {\em safe set} ${\C \subset \X \times \R}$, stated as follows.
\begin{definition}
System~\eqref{eq:closedloop} is called {\em safe w.r.t.~$\C$} if ${(\xx(t),t) \!\in\! \C}$, ${\forall t \!\geq\! 0}$ holds for all ${\xx_0 \!\in\! \X}$ that satisfies ${(\xx_0,0) \!\in\! \C}$.
In other words, safety means that set $\C$ is forward invariant along~\eqref{eq:closedloop}.
\end{definition}
Let $\C$ be given as the 0-superlevel set of a continuously differentiable function ${h : \X \times \R \to \R}$, with boundary ${\partial \C}$:
\begin{align}
\begin{split}
    \C & = \{(\xx,t) \in \X \times \R: h(\xx,t) \geq 0 \}, \\
    \partial \C & = \{(\xx,t) \in \X \times \R: h(\xx,t) = 0 \}.
\end{split}
\label{eq:safeset}
\end{align}
Then, we characterize safety using the scalar-valued function $h$, whose positive (or negative) sign indicates safe (or unsafe) behavior.
In the context of aircraft collision avoidance or geofencing, $h$ is related to the signed distance of the aircraft from other aircraft or the geofence, respectively, that must kept positive for safety.
The expression of $h$ will be detailed below.
Throughout the paper, we assume that the underlying set $\C$ is non-empty, it has no isolated points, and that $\derp{h}{\xx}(\xx,t)$ is nonzero for all ${(\xx,\!t) \!\in\! \partial \C}$ (i.e., 0 is a regular value of $h$).

Given set $\C$ and function $h$, the theory of CBFs establishes a method to synthesize controllers with formal safety guarantees.
CBFs are defined below, by using the derivative of $h$ along system~\eqref{eq:system} that quantifies the effect of the input $\uu$ on safety:
\begin{equation}
    \dot{h}(\xx,t,\uu) =
    \derp{h}{t}(\xx,t) \!+\!
    \derp{h}{\xx}(\xx,t) \ff(\xx) \!+\!
    \derp{h}{\xx}(\xx,t) \gg(\xx)
    \uu.
\label{eq:hdot}
\end{equation}

\begin{definition}[\cite{AmesXuGriTab2017}] \label{def:CBF}
Function $h$ is a {\em control barrier function} for~\eqref{eq:system} on $\C$ if there exists ${\alpha \!\in\! \Keinf}$ such that for all ${(\xx,t) \!\in\! \C}$:
\begin{equation}
    \sup_{\uu \in \U} \dot{h}(\xx,t,\uu) > - \alpha \big( h(\xx,t) \big).
\label{eq:CBF_condition}
\end{equation}
\end{definition}

Given a CBF, the following theorem establishes formal safety guarantees for controllers.
\begin{lemma}[\cite{AmesXuGriTab2017}] \label{lemma:CBF}
\textit{
If $h$ is a CBF for~\eqref{eq:system} on $\C$, then any locally Lipschitz continuous controller $\kk$ that satisfies: 
\begin{equation}
    \dot{h} \big( \xx, t, \kk(\xx,t) \big) \geq - \alpha \big( h(\xx,t) \big)
\label{eq:safety_condition}
\end{equation}
for all ${(\xx,t) \in \C}$ renders~\eqref{eq:closedloop} safe w.r.t.~$\C$.
}
\end{lemma}

Note that if the derivative of $h$ is independent of the control input $\uu$, i.e., ${\derp{h}{\xx}(\xx,t) \gg(\xx) \equiv \mathbf{0}}$ in~\eqref{eq:hdot}, then $h$ cannot be used directly to synthesize safe controllers.
In such cases, methods like high-order CBFs~\cite{Nguyen2016, xiao2019cbf} or backstepping~\cite{taylor2022safebackstepping}  can be applied to construct a CBF from $h$ and use it for control.
These methods are discussed further in Section~\ref{sec:RTA}.
The main idea behind high-order CBFs is introducing ${\dot{h}+\alpha(h)}$, that occurs in~\eqref{eq:safety_condition}, as a high-order (or extended) CBF:
\begin{equation}
    h_{\rm e}(\xx,t) =
    \derp{h}{t}(\xx,t) +
    \derp{h}{\xx}(\xx,t) \ff(\xx) +
    \alpha \big( h(\xx,t) \big),
\label{eq:hocbf}
\end{equation}
whose 0-superlevel set is $\C_{\rm e}$, cf.~\eqref{eq:safeset}.
If the extended CBF $h_{\rm e}$ is maintained nonnegative, then safety can be guaranteed based on~\eqref{eq:safety_condition}, similar to the case when $h$ is a CBF.
\begin{lemma}[\cite{xiao2019cbf}] \label{lemma:HOCBF}
\textit{
If $h_{\rm e}$ is a CBF for~\eqref{eq:system} on $\C_{\rm e}$, then any locally Lipschitz continuous controller $\kk$ that satisfies: 
\begin{equation}
    \dot{h}_{\rm e} \big( \xx, t, \kk(\xx,t) \big) \geq - \alpha \big( h_{\rm e}(\xx,t) \big)
\label{eq:safety_condition_hocbf}
\end{equation}
for all ${(\xx,t) \in \C \cap \C_{\rm e}}$ renders~\eqref{eq:closedloop} safe w.r.t.~$\C \cap \C_{\rm e}$.
}
\end{lemma}

\subsection{Safety Filters}

Based on Lemma~\ref{lemma:CBF}, controllers can be designed for~\eqref{eq:system} in a safety-critical fashion by enforcing~\eqref{eq:safety_condition} during controller synthesis.
Such safety-critical control designs are often established by so-called {\em safety filters} that modify a desired but not necessarily safe controller ${\kd : \X \times \R \to \U}$ to a safe controller $\kk$.
In the context of aircraft control, the desired controller $\kd$ may be a flight controller that tracks a desired trajectory, such as the controller in Appendix~\ref{appdx:velocity_tracking}.
This can be modified for safety, for example, by the optimization problem:
\begin{align}
\begin{split}
    \kk(\xx,t) =
    \underset{\uu \in \U}{\operatorname{argmin}} & \quad
    \big\| \uu - \kd(\xx,t) \big\|_{\GGamma}^2 \\
    \text{s.t.} & \quad
    \dot{h}(\xx,t,\uu) \geq - \alpha \big( h(\xx,t) \big),
\end{split}
\label{eq:QP}
\end{align}
that minimizes the difference of the desired and actual inputs subject to the safety constraint~\eqref{eq:safety_condition}.
Note that a symmetric positive definite weight matrix ${\GGamma \in \R^{m \times m}}$ is useful for scaling the various components of the control input when these have different physical meanings and orders of magnitude.

We remark that the formulation above does not include input constraints (as ${\uu \in \U}$).
To enforce input limits, one could saturate the controller, or add input constraints to~\eqref{eq:QP} and relax the safety constraint to maintain feasibility.
However, these methods lose formal safety guarantees.
Formally addressing input constraints while maintaining safety requires more advanced CBF approaches, such as those in~\cite{gurriet2020scalable, agrawal2021safe, Ames2021icbf, liu2022safe, xiao2022sufficient}.
These methods are beyond the scope of this paper and will be explored as future work.
Instead, the controllers proposed here are tuned so that the resulting control inputs remain within reasonable bounds in simulations.

Importantly, the optimization problem~\eqref{eq:QP} is feasible if $h$ is indeed a CBF that satisfies~\eqref{eq:CBF_condition}, and it can be solved in closed form.
By factorizing $\GGamma$ as ${\GGamma = \WW^{-\top} \WW^{-1}}$ with positive definite ${\WW \in \R^{m \times m}}$,~\eqref{eq:QP} transforms into:
\begin{align}
\begin{split}
    & \hspace{-26mm} \kk(\xx,t) = \kd(\xx,t) + \WW \hat{\kk}(\xx,t), \\
    \hat{\kk}(\xx,t) =
    \underset{\hat{\uu} \in \U}{\operatorname{argmin}} & \quad
    \|\hat{\uu}\|^2 \\
    \text{s.t.} & \quad
    a(\xx,t) + \bb(\xx,t) \hat{\uu} \geq 0,
\end{split}
\label{eq:QP_transformed}
\end{align}
where:
\begin{align}
\begin{split}
    a(\xx,t) & = \dot{h} \big( \xx, t, \kd(\xx,t) \big) + \alpha \big( h(\xx,t) \big), \\
    \bb(\xx,t) & = \derp{h}{\xx}(\xx,t) \mb{g}(\xx) \WW.
\end{split}
\end{align}
Then~\eqref{eq:QP_transformed} can be solved as follows~\cite{cohen2023smooth}:
\begin{equation}
    \kk(\xx,t) = \kd(\xx,t) + \Lambda \big( a(\xx,t), \| \bb(\xx,t) \| \big) \WW \bb(\xx,t)^\top,
\label{eq:safetyfilter}
\end{equation}
with:
\begin{equation}
    \Lambda(a,b) =
    \begin{cases}
        0 & {\rm if}\ b = 0, \\
        \frac{1}{b} \max \Big\{ 0, - \frac{a}{b} \Big\} & {\rm if}\ b \neq 0.
    \end{cases}
\label{eq:safetyfilter_max}
\end{equation}
Since the safety filter~\eqref{eq:safetyfilter} is given explicitly, it is fast and easy to evaluate it, which is an advantage of using CBFs.

Finally, we remark that the controller~\eqref{eq:safetyfilter}-\eqref{eq:safetyfilter_max} is continuous but not necessarily differentiable.
Yet, differentiable controllers may be preferable in certain scenarios.
As established by~\cite{cohen2023smooth}, a smooth over-approximation of $\Lambda$ in~\eqref{eq:safetyfilter_max}, such as:
\begin{equation}
    \Lambda(a,b) =
    \begin{cases}
        0 & {\rm if}\ b = 0, \\
        \frac{1}{\nu b} \ln \Big( 1 + {\rm e}^{- \nu \frac{a}{b}} \Big) & {\rm if}\ b \neq 0,
    \end{cases}
\label{eq:safetyfilter_smooth}
\end{equation}
with parameter ${\nu > 0}$, leads to a smooth controller in~\eqref{eq:safetyfilter} that still satisfies~\eqref{eq:safety_condition} (and it approaches~\eqref{eq:safetyfilter_max} for ${\nu \to \infty}$).
Henceforth, we will use the safety filter~\eqref{eq:safetyfilter}-\eqref{eq:safetyfilter_max} and the smooth safety filter~\eqref{eq:safetyfilter}-\eqref{eq:safetyfilter_smooth} for safety-critical control.

\subsection{Compositions of Safety Constraints}

In practice, control systems may need to satisfy more than one safety constraints simultaneously.
In such cases, the safe set $\C$ is composed of multiple sets $\C_{i}$ related to each safety constraint.
For example, as illustrated in Fig.~\ref{fig:concept}, an aircraft may execute both collision avoidance and geofencing at the same time, associated with sets $\C_{\rm coll}$ and $\C_{\rm geo}$.
This leads to the overall safe set ${\C = \C_{\rm coll} \cap \C_{\rm geo}}$.
With this as motivation, now we briefly revisit a method established in~\cite{molnar2023composing} to combine multiple safety constraints and express them by a single CBF.

First, consider the scenario when safety must be maintained against $N_{\rm c}$ constraints simultaneously, associated with sets $\C_{i}$, functions $h_{i}$, and index ${i \in I = \{1, \ldots, N_{\rm c} \}}$.
Safety is interpreted w.r.t.~the intersection ${\C = \bigcap_{i \in I} \C_{i}}$ of the sets:
\begin{equation}
    \bigcap_{i \in I} \C_{i} = \Big\{ (\xx,t) \in \X \times \R: \min_{i \in I} h_{i}(\xx,t) \geq 0 \Big\}.
\label{eq:mincbf}
\end{equation}
Then, a single CBF candidate can be constructed by a smooth approximation of the $\min$ function~\cite{molnar2023composing, lindemann2019stl}:
\begin{equation}
    h(\xx,t) = - \frac{1}{\kappa} \ln \bigg( \sum_{i \in I} {\rm e}^{- \kappa h_{i}(\xx,t)} \bigg),
\label{eq:hmin}
\end{equation}
with a smoothing parameter ${\kappa>0}$, where the approximation error is bounded by:
\begin{equation}
    - \frac{\ln N_{\rm c}}{\kappa} \leq h(\xx,t) - \min_{i \in I} h_{i}(\xx,t) \leq 0,
\label{eq:hmin_bounds}
\end{equation}
and diminishes as ${\kappa \to \infty}$.
Formula~\eqref{eq:hmin} establishes a simple way to combine multiple barriers, ultimately providing a single CBF candidate for use in the safety filter~\eqref{eq:safetyfilter}, with derivative:
\begin{equation}
    \dot{h}(\xx,t,\uu) = \sum_{i = 1}^{N_{\rm c}} {\rm e}^{-\kappa (h_{i}(\xx,t) - h(\xx,t))} \dot{h}_{i}(\xx,t,\uu).
\end{equation}

Formula~\eqref{eq:hmin} includes the evaluation of $N_{\rm c}$ exponential functions, hence its computation time increases with the number $N_{\rm c}$ of constraints.
Yet, a significant benefit of constructing a single CBF by~\eqref{eq:hmin} is that it enables the use of closed-form safety filters like~\eqref{eq:safetyfilter} that are easy to compute.
Without such composition, safety filters would require solving optimization problems like~\eqref{eq:QP} with $N_{\rm c}$ number of constraints, which would ultimately lead to more computation especially if $N_{\rm c}$ is large.
Further details on the properties of the CBF composition method in~\eqref{eq:hmin} can be found in~\cite{molnar2023composing}.

The framework in~\cite{molnar2023composing} can also be used for more complex compositions of safe sets, as combinations of set intersections and unions, which correspond to AND and OR logic between safety constraints, respectively.
For example, the geofence in Fig.~\ref{fig:concept} can also be expressed by unions and intersections: ${\C_{\rm geo} = \C_{1} \cup \C_{2} \cap \C_{3}}$.
As such, the union of individual sets:
\begin{equation}
    \bigcup_{i \in I} \C_{i} = \Big\{ (\xx,t) \in \X \times \R: \max_{i \in I} h_{i}(\xx,t) \geq 0 \Big\}
\label{eq:maxcbf}
\end{equation}
can be captured by a single CBF candidate analogously to~\eqref{eq:hmin}:
\begin{equation}
    h(\xx,t) = \frac{1}{\kappa} \ln \bigg( \sum_{i \in I} {\rm e}^{\kappa h_{i}(\xx,t)} \bigg).
\label{eq:hmax}
\end{equation}
Combinations of set unions and intersections can be addressed by the recursive applications of formulas~\eqref{eq:hmin} and~\eqref{eq:hmax}.
Note that~\eqref{eq:CBF_condition} must hold for $h$ to be a valid CBF.
This requires a control-sharing property~\cite{xu2018constrained, katriniok2022controlsharing} that the individual barriers $h_{i}$ are compatible and do not lead to contradicting constraints for the control input.
Establishing the conditions for obtaining a valid CBF after a general composition is still an open problem, and initial results can be found in~\cite{molnar2023composing}.
Overall, formulas~\eqref{eq:hmin} and~\eqref{eq:hmax} provide a systematic way of encoding complex compositions of safety constraints into a single CBF candidate.
This will be leveraged below in the context of simultaneous collision avoidance and geofencing where the geofence is composed of multiple boundaries.

\section{Run-time Assurance on Fixed-wing Aircraft}
\label{sec:RTA}

Now we present our main contributions, wherein we use CBFs to formally guarantee safety on fixed-wing aircraft in collision avoidance and geofencing tasks.
We establish safety-critical controllers for the system~\eqref{eq:Dubins_affine} by utilizing the safety filter~\eqref{eq:safetyfilter} as run-time assurance (RTA) system with an appropriate choice of CBF.
We address the nontrivial problem of CBF synthesis for aircraft dynamics, discuss various choices of CBFs, and demonstrate the performance of the resulting controllers by numerical simulations.

\subsection{Position-based CBF Candidates}

We seek to execute simultaneous collision avoidance and geofencing on the aircraft in a provably safe fashion via CBFs.
Both collision avoidance and geofencing correspond to safety constraints on the position $\rr$ of the aircraft.
Accordingly, we have a {\em safe set} $\S$ defined over the position space and time:
\begin{equation}
    \S = \{(\rr,t) \in \R^3 \times \R_{\geq 0} : \hp(\rr,t) \geq 0 \},
\end{equation}
where $\hp$ is a position-based CBF candidate described below.
We seek to ensure ${(\rr(t),t) \in \S}$, ${\forall t \geq 0}$ for all ${(\rr_0,0) \in \S}$.
Note that while we focus on position constraints only, attitude constraints could be addressed in a similar fashion by constructing CBFs that depend on the Euler angles.

\subsubsection{Collision avoidance}

To avoid collisions, we enforce that the signed distance between the controlled aircraft and another aircraft is nonnegative for all time.
That is, we require
${\hpi(\rr(t),t) \geq 0}$, ${\forall t \geq 0}$,
with:
\begin{equation}
    \hpi(\rr,t) = \| \rr - \rr_{i}(t) \| - \rho_{i},
\label{eq:CBF_coll}
\end{equation}
where $\rr_{i}(t)$ is the other aircraft's position, and  ${\rho_{i} > 0}$ is a collision radius containing the combined size of both aircraft and an additional buffer distance if desired.
If there are multiple other aircraft, we get multiple collision constraints with index $i$.
The corresponding derivatives are:
\begin{equation}
    \dot{h}_{{\rm p},i}(\rr,t,\vv) =  \nn_{i}(\rr,t)^\top \big( \vv - \vv_{i}(t) \big),
\end{equation}
with ${\dot{\rr}_{i}(t) = \vv_{i}(t)}$ and:
\begin{equation}
    \nn_{i}(\rr,t) = \frac{\rr - \rr_{i}(t)}{\| \rr - \rr_{i}(t) \|}.
\end{equation}

\subsubsection{Geofencing}

For geofencing, we keep the aircraft on one side of a ``fence'', modeled as a plane at $\rr_{i}$ with unit normal vector $\nn_{i}$.
That is, we require
${\hpi(\rr(t),t) \geq 0}$, ${\forall t \geq 0}$, where:
\begin{equation}
    \hpi(\rr,t) = \nn_{i}^\top (\rr - \rr_{i}) - \rho_{i},
\label{eq:CBF_geo}
\end{equation}
and ${\rho_{i} \geq 0}$ is the distance to be kept from the geofence.
Again, index $i$ indicates the possibility of multiple geofence constraints, that is, geofences with more complex geometry.
The derivatives of the CBF candidates are:
\begin{equation}
    \dot{h}_{{\rm p},i}(\rr,t,\vv) = \nn_{i}^\top \vv.
\end{equation}
Note that zero is a regular value of $\hpi$ for both~\eqref{eq:CBF_coll} and~\eqref{eq:CBF_geo}, i.e., the gradient of $\hpi$ is nonzero when ${\hpi(\rr,t)=0}$.

\subsubsection{Simultaneous collision avoidance and geofencing}

Given $N_{\rm c}$ number of safety constraints, these can be combined into a single CBF candidate $\hp(\rr(t),t)$ based on formulas~\eqref{eq:hmin} and~\eqref{eq:hmax}.
For simplicity, we make the following assumption.
\begin{assumption} \label{assum:composition}
Safety constraints are linked with AND logic: one must avoid collision with aircraft~1 AND aircraft~2, etc., AND obey geofence~1 AND geofence~2, and so on.
\end{assumption}
\noindent This assumption is made only to keep the exposition simple, since AND logic only requires~\eqref{eq:hmin} for CBF composition.
Combination of AND and OR logic could also be handled straightforwardly via combining~\eqref{eq:hmin} with~\eqref{eq:hmax}.
Considering Assumption~\ref{assum:composition} and using~\eqref{eq:hmin} leads to the CBF candidate:
\begin{equation}
    \hp(\rr,t) = -\frac{1}{\kappa} \ln \bigg( \sum_{i = 1}^{N_{\rm c}} {\rm e}^{-\kappa \hpi(\rr,t)} \bigg),
\label{eq:logsumexp}
\end{equation}
with ${\kappa > 0}$.
Note that ${\hp(\rr,t) = \hpi(\rr,t)}$ if ${N_{\rm c} = 1}$.
We remark that~\eqref{eq:logsumexp} approximates a convex polytope for geofencing when $\hpi$ are given by~\eqref{eq:CBF_geo}.
Non-convex regions could be obtained as the union of convex polytopes, i.e., by using~\eqref{eq:hmax}.

Ideally, we would use the position-based CBF candidate $\hp$ directly as CBF:  ${h(\xx,t)=\hp(\rr,t)}$.
However, the corresponding safety constraint in~\eqref{eq:safety_condition} with ${\alpha_{\rm p} \in \Keinf}$:
\begin{equation}
    \dot{h}_{\rm p} \big( \rr, t, \vv(\zzeta) \big) \geq - \alpha_{\rm p} \big( \hp(\rr,t) \big)
\label{eq:safety_condition_velocity}
\end{equation}
is independent of the controller and may not always hold.
As a result,~\eqref{eq:CBF_condition} in Definition~\ref{def:CBF} is not satisfied, $\hp$ is not a CBF, and $\hp$ cannot be used directly to synthesize safe controllers for~\eqref{eq:Dubins_affine}.
We rather call $\hp$ as {\em CBF candidate}, and below we use it to construct a CBF $h$ for controller synthesis.

\subsection{Run-time Assurance with Velocity-based Extended CBF}
\label{sec:extended}

\bgroup
\setlength{\tabcolsep}{3pt}
\begin{table}
\caption{Parameter Values for Numerical Simulations (with SI Units)}
\begin{center}
\begin{tabular}{c|ccc}
model &
${\gD \!=\! 9.81}$ & & \\ 
\hline
safety filter &
${\alpha(r) \!=\! \gamma r}$ &
${\gamma \!=\! 0.1}$ & 
${\WW \!=\! \begin{bmatrix}
6 \!\!&\!\! 0 \!\!&\!\! 0 \\
0 \!\!&\!\! 0.6 \!\!&\!\! 0 \\ 0 \!\!&\!\! 0 \!\!&\!\! 0.1 \end{bmatrix}}$ \\ 
\hline
coll. avoid &
${\rr_{1}(t) \!=\! \mb{c} \!+\! \vv_{1} t}$ &
$\mb{c} \!=\! \begin{bmatrix} -3048 \\ 0 \\ 0 \end{bmatrix}$ & 
$\vv_{1} \!=\! \begin{bmatrix} 121.92 \\ 161.32 \\ 0 \end{bmatrix}$ \\ 
\hline
geofence &
${\rr_{2,3} \!=\! \begin{bmatrix} 0 \\ 11901 \\ 0 \end{bmatrix}}$ & 
${\nn_{2} \!=\! \dfrac{1}{\sqrt{17}} \begin{bmatrix} -4 \\ -1 \\ 0 \end{bmatrix}}$ &
${\nn_{3} \!=\! \dfrac{1}{\sqrt{5}} \begin{bmatrix} -2 \\ -1 \\ 0 \end{bmatrix}}$ \\
\hline
pos. CBF &
${\rho_{1} \!=\! 30}$ & 
${\rho_{2,3} \!=\! 15}$ & 
${\kappa \!=\! 0.007}$ 
\\
\hline
ext. CBF &
${\gamma_{\rm p} \!=\! 0.1}$ & 
&
\\
\hline
backstepping &
${\alpha_{\rm e}(r) \!=\! \gamma_{\rm e} r}$ &
${\gamma_{\rm e} \!=\! 0.1}$ & 
${\WW_{\rm e} \!=\! \II}$ \\ 
CBF &
${\nu_{\rm e} \!=\! 1}$ &
${\mu_{\rm e} \!=\! 10^{-4}}$ & 
\\
\hline
model-free &
${\alpha_{\rm p}(r) \!=\! \gamma_{\rm p} r}$ &
${\gamma_{\rm p} \!=\! 0.1}$ & 
${\sigma \!=\! 3}$ \\ 
RTA &
${\Gamma_{\vv} \!=\! 4}$ &
${\nu_{\vv} \!=\! 0.007}$ & 
\\
\hline
\multirow{2}{*}{traj. tracking} &
${\rr_{\rm g}(t) \!=\! \vv_{\rm g} t}$ &
${\vv_{\rm g} \!=\! \begin{bmatrix} 0 \\ 161.32 \\ 0 \end{bmatrix}}$ & 
${\KK_{\rr} \!=\! 0.05\, \II}$ \\ 
 &
${\KK_{\vv} \!=\! 0.3\, \II}$ & 
${\mu \!=\! 10^{-5}}$ & 
${\lambda \!=\! 0.2}$ \\ 
\end{tabular}
\end{center}
\label{tab:parameters}
\end{table}
\egroup

First, we employ so-called {\em extended CBFs} (or high-order CBFs) proposed in~\cite{Nguyen2016, xiao2019cbf}.
Extended CBFs, defined as follows, depend on the velocity $\vv$:
\begin{equation}
    \hei(\rr,\vv,t) = \hpi(\rr,t) + \frac{1}{\gamma_{\rm p}} \dot{h}_{{\rm p},i} \big( \rr, t, \vv \big),
\label{eq:extended_CBF_def}
\end{equation}
with ${\gamma_{\rm p}>0}$.
For collision avoidance, the extended CBF is:
\begin{equation}
    \hei(\rr,\vv,t) = \| \rr - \rr_{i}(t) \| - \rho_{i} + \frac{1}{\gamma_{\rm p}} \nn_{i}(\rr,t)^\top \big( \vv - \vv_{i}(t) \big),
\label{eq:extended_CBF_coll}
\end{equation}
cf.~\eqref{eq:CBF_coll}.
For geofencing, the extended CBF becomes:
\begin{equation}
    \hei(\rr,\vv,t) = \nn_{i}^\top (\rr - \rr_{i}) - \rho_{i} + \frac{1}{\gamma_{\rm p}} \nn_{i}^\top \vv,
\label{eq:extended_CBF_geo}
\end{equation}
cf.~\eqref{eq:CBF_geo}.
It can be shown that zero is a regular value of $\hei$ in~\eqref{eq:extended_CBF_coll}-\eqref{eq:extended_CBF_geo}, i.e., the gradient of $\hei$ is nonzero when ${\hei(\rr,\vv,t)=0}$.
For simultaneous collision avoidance and geofencing, the extended CBFs can be combined as in~\eqref{eq:hmin}:
\begin{equation}
    \he(\rr,\vv,t) = -\frac{1}{\kappa} \ln \bigg( \sum_{i = 1}^{N_{\rm c}} {\rm e}^{-\kappa \hei(\rr,\vv,t)} \bigg).
\label{eq:logsumexp_extended}
\end{equation}

\begin{figure}[!t]
\centering
\includegraphics[scale=1]{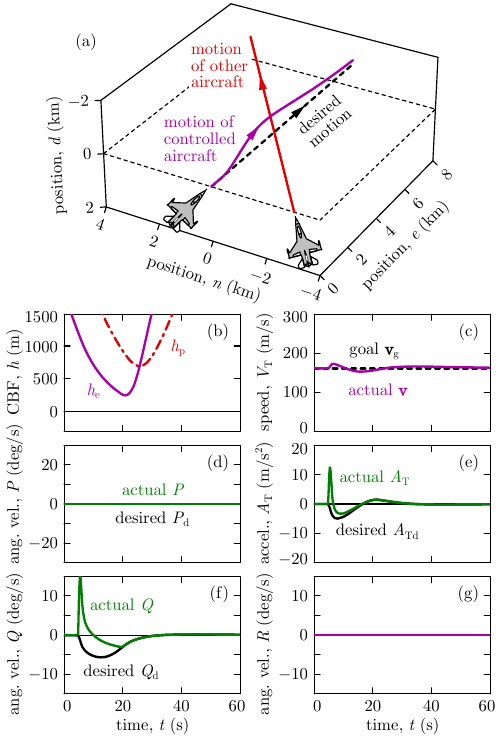}
\caption{Simulation of RTA in collision avoidance using extended CBF.
The aircraft successfully maintains safety.}
\label{fig:sim_extended_coll}
\end{figure}

With these definitions, we use the extended CBFs directly:
\begin{equation}
    h(\xx,t) = \he \big( \rr,\vv(\zzeta),t \big).
\label{eq:extended_CBF}  
\end{equation}
Importantly, enforcing ${h(\xx,t) \geq 0}$ implies that~\eqref{eq:safety_condition_velocity} holds with ${\alpha_{\rm p}(r) = \gamma_{\rm p} r}$.
Therefore, based on applying Lemma~\ref{lemma:HOCBF}, the safety filter~\eqref{eq:safetyfilter} guarantees ${h(\xx(t),t) \geq 0}$, ${t \geq 0}$ and ${\hp(\rr(t),t) \geq 0}$, ${\forall t \geq 0}$ (if ${h(\xx_0,0) \geq 0}$ and ${\hp(\rr_0,0) \geq 0}$ hold).
This means guaranteed safety in collision avoidance and geofencing.
Details and proof of the safety guarantees provided by extended CBFs are in~\cite{xiao2019cbf}.

The performance of the proposed safety-critical controller is demonstrated by simulating the aircraft model~\eqref{eq:Dubins_affine} with the safety filter~\eqref{eq:safetyfilter}-\eqref{eq:safetyfilter_max} and the extended CBF~\eqref{eq:extended_CBF_def}-\eqref{eq:extended_CBF}.
The simulation parameters used throughout the paper are listed in Table~\ref{tab:parameters}, and the desired controller $\kd$ is considered to be the trajectory tracking controller detailed in Appendix~\ref{appdx:velocity_tracking}.
Note that safety guarantees do not depend on the choice of desired controller, other flight controllers could also be used in RTA.

Fig.~\ref{fig:sim_extended_coll} illustrates a collision avoidance scenario.
First, the controlled aircraft is tracking a straight-line trajectory at constant speed with its desired controller $\kd$.
Then, another aircraft approaches from the right, which would result in a collision if the controlled aircraft did not respond.
Thus, the safety filter intervenes and the controller $\kk$ starts to deviate from the desired one $\kd$.
As a result, the controlled aircraft accelerates while pitching and moving up to safely avoid the other aircraft.
Then, the safety filter stops intervening, and the aircraft starts to use its desired controller $\kd$ to track its original course once again.
Remarkably, the collision avoidance maneuver is generated automatically by the CBF.

Observe in Fig.~\ref{fig:sim_extended_coll}(d)-(g), however, that the aircraft only leverages acceleration and pitching for collision avoidance, while it refrains from turning left or right via rolling and yawing.
The safety filter only modifies two of the control inputs, $\AT$ and $Q$, that cause acceleration and pitching, while it does not affect the third input, $P$, that would induce rolling and consequently turning.
This behavior is caused by the cascaded structure~\eqref{eq:Dubins_cascade} of the dynamics and the construction of the extended CBF.
Namely, the inputs $\AT$ and $Q$ enter the cascaded dynamics at the 2nd level through the evolution of the velocity-related states $\zzeta$, while the input $P$ shows up at the 3rd level in the equation of the roll angle $\phi$.
Since the velocity-dependent extended CBF $\he \big( \rr, \vv(\zzeta), t \big)$ includes the states $\zzeta$ but not the roll angle $\phi$, the resulting safety filter includes $\AT$ and $Q$ but not $P$.
Hence, the safety filter cannot make the plane roll, which prevents it from turning left or right; cf.~\eqref{eq:Dubins_R}.

\begin{figure}[!t]
\centering
\includegraphics[scale=1]{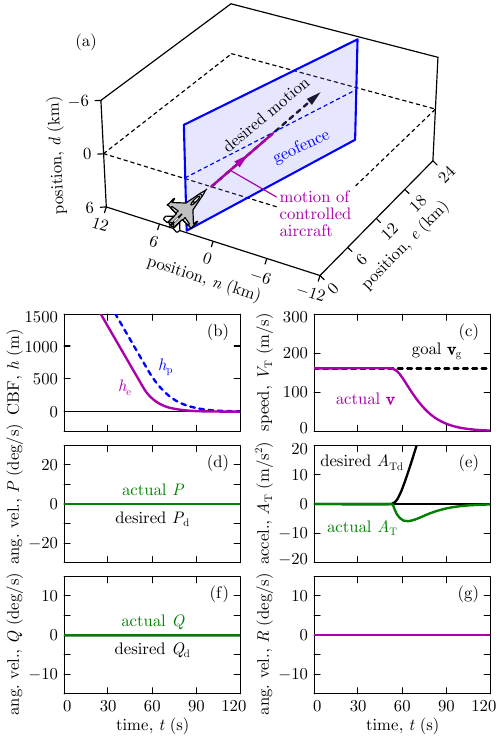}
\caption{Simulation of RTA in geofencing using extended CBF.
The controller fails this task as it would command the aircraft to stop in front of the geofence.
This CBF choice lacks the ability to make the aircraft turn left or right.}
\label{fig:sim_extended_geo}
\end{figure}

While the collision avoidance example was successful, lacking the ability to turn can be detrimental for RTA.
Fig.~\ref{fig:sim_extended_geo} highlights this by showcasing a geofencing scenario with a vertical plane as geofence.
Since the extended CBF $\he$ cannot induce turning, the safety filter forces the aircraft to slow down and stop in front of the geofence.
Although this behavior is safe from geofencing point of view, it is obviously infeasible in practice to make the aircraft stop.
The geofencing task could only be accomplished with the ability to turn.
This motivates us to construct a better CBF, which respects the cascaded structure of the dynamics and incorporates all states (including the roll angle $\phi$) so that the safety filter leverages all control inputs (including the roll motion $P$).

\subsection{Run-time Assurance with Backstepping-based CBF}
\label{sec:backstepping}

While the extended CBF-based RTA does not leverage all possible behaviors (acceleration, pitching, rolling, and yawing), this can be done by other CBF choices~\cite{Xiao2022}.
In particular, the method of {\em backstepping}~\cite{taylor2022safebackstepping} offers a systematic procedure to synthesize valid CBFs for cascaded systems like~\eqref{eq:Dubins_cascade}.
Here we use backstepping to construct a CBF for use in RTA rather than construct the control law itself.
Importantly, backstepping is able to provide a valid CBF whose derivative is affected by all control inputs, hence the underlying RTA can leverage all aforementioned behaviors for safety---even considering 3D motions and nontrivial cascaded dynamics.

We proceed with backstepping to design a CBF in two steps:
\begin{enumerate}
    \item We apply the extended CBF to design a safe acceleration $\as$ and a corresponding safe angular velocity $\Rs$ (related to turning).
    Note, however, that the aircraft cannot be commanded directly to turn, as $R$ is not an input.
    \item We use the safe angular velocity $\Rs$ to construct a CBF based on backstepping. This will allow us to synthesize the remaining input, the angular velocity $P$ (related to rolling), along with the other two inputs $\AT$ and $Q$.
\end{enumerate}

To synthesize the safe angular velocity $\Rs$, we first consider the extended CBF-based safety constraint:
\begin{equation}
    \dot{h}_{\rm e}(\rr,\vv,t,\aa) \geq - \alpha_{\rm e} \big( \he(\rr,\vv,t) \big),
\end{equation}
and the corresponding smooth safety filter to obtain the safe acceleration $\as$ (associated with zero desired acceleration):
\begin{align}
\begin{split}
    \as(\xx,t) & =
    \Lambda \big( a_{\rm e}(\xx,t), \| \bb_{\rm e}(\xx,t) \| \big) \WW_{\rm e} \bb_{\rm e}(\xx,t)^\top, \\
    a_{\rm e}(\xx,t) & = \dot{h}_{\rm e} \big( \rr,\vv(\zzeta),t,\OO \big) + \alpha_{\rm e} \big( \he \big( \rr,\vv(\zzeta),t \big) \big), \\
    \bb_{\rm e}(\xx,t) & = \derp{\he}{\vv} \big( \rr,\vv(\zzeta),t \big) \WW_{\rm e},
\end{split}
\end{align}
with $\Lambda$ in~\eqref{eq:safetyfilter_smooth}, a smoothing parameter ${\nu_{\rm e}>0}$, a weight $\WW_{\rm e}$ on the various acceleration components, and ${\alpha_{\rm e} \in \Keinf}$.
We convert the safe acceleration $\as$ to the angular velocity $\Rs$ based on the relation~\eqref{eq:Dubins_acceleration} listed in Appendix~\ref{appdx:model}:
\begin{equation}
    \Rs = \WW_{R}(\zzeta,\phi) \as(\xx,t),
\label{eq:Dubins_R_conversion}
\end{equation}
where $\WW_{R}$ is the last row of the inverse of $\MM_{\aa}$ in~\eqref{eq:Dubins_acceleration_M}.
Note that the use of a smooth safety filter makes $\Rs$ differentiable.

Finally, with the safe angular velocity $\Rs$, we construct the following {\em backstepping-based CBF} proposed by~\cite{taylor2022safebackstepping}:
\begin{equation}
    \hb(\xx,t) = \he \big( \rr,\vv(\zzeta),t \big) - \frac{1}{2 \mu_{\rm e}} (\Rs - R)^2,
\label{eq:Dubins_CBF_backstepping}
\end{equation}
with a scaling constant ${\mu_{\rm e} > 0}$.
Here $R$ and $\Rs$ are given by~\eqref{eq:Dubins_R} and~\eqref{eq:Dubins_R_conversion}.
Notice that ${\hb(\xx,t) \leq \he \big( \rr,\vv(\zzeta),t \big)}$ holds, hence safety w.r.t.~$\hb$ implies safety w.r.t.~$\he$.
The CBF ${h(\xx,t)=\hb(\xx,t)}$ can be used to execute the safety filter~\eqref{eq:safetyfilter} for RTA, where the derivative of $\hb$ can be obtained through lengthy calculation by differentiating the expression of $\Rs$.
Incorporating the turning rate $R$ into the CBF $\hb$ allows us to leverage all control inputs for safety, including the angular velocity $P$ required for making the aircraft roll and consequently turn.
This process means backstepping: taking a step back from rolling ($P$) to turning ($R$).

The backstepping-based CBF can be established for collision avoidance (with $\hei$ from~\eqref{eq:extended_CBF_coll}), geofencing (with $\hei$ from~\eqref{eq:extended_CBF_geo}), and the combination thereof (with $\he$ from~\eqref{eq:logsumexp_extended}).
Since the backstepping calculations are nontrivial, it is advantageous to first combine multiple safety constraints into a single one before executing backstepping rather than doing backstepping multiple times and then combining.
The resulting safety filter~\eqref{eq:safetyfilter} is in closed form.
It yields safety w.r.t.~the backstepping-based CBF, ${\hb(\xx(t),t) \geq 0}$, ${\forall t \geq 0}$, which implies safety w.r.t.~the extended CBF, ${\he \big( \rr(t),\vv(\zzeta(t)),t \big) \geq 0}$, ${\forall t \geq 0}$, which finally leads to safe behavior considering the position-based CBF candidate, ${\hp(\rr(t),t) \geq 0}$, ${\forall t \geq 0}$.
Note that the backstepping-based CBF~\eqref{eq:Dubins_CBF_backstepping} contains an additional term compared to the extended CBF~\eqref{eq:extended_CBF_def}, which is obtained by adding a term to the position-based CBF candidate.
Thus, these CBFs have different numbers of parameters.
To compare these CBFs and the resulting controllers, we use the same values for the parameters shared by the various CBFs (see~Table~\ref{tab:parameters}), and the comparison reflects the effect of the added terms.

\begin{figure}[!t]
\centering
\includegraphics[scale=1]{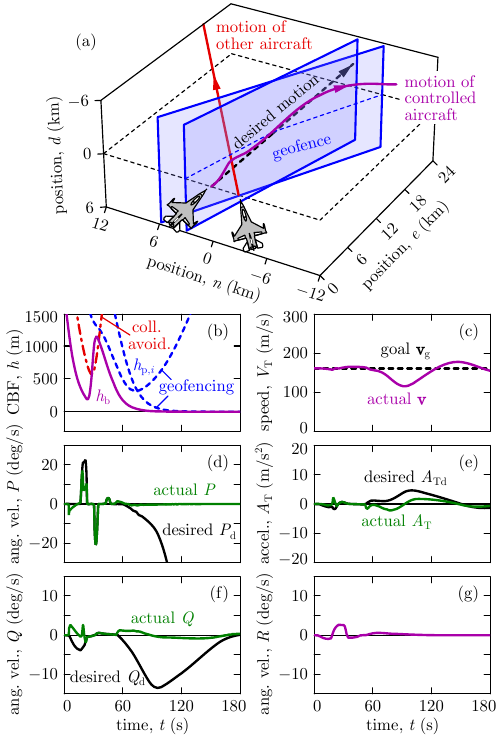}
\caption{Simulation of RTA in simultaneous collision avoidance and geofencing using backstepping-based CBF.
Safety is successfully maintained by leveraging both acceleration, pitching, rolling, and turning.}
\label{fig:sim_backstepping}
\end{figure}

The behavior of the aircraft with the proposed safety-critical controller is showcased in Fig.~\ref{fig:sim_backstepping}.
Model~\eqref{eq:Dubins_affine} is simulated with the safety filter~\eqref{eq:safetyfilter}-\eqref{eq:safetyfilter_max}, the backstepping-based CBF~\eqref{eq:Dubins_CBF_backstepping}, and the parameters in Table~\ref{tab:parameters}.
The simulated scenario is simultaneous collision avoidance and geofencing where the geofence consists of two planar boundaries.
The aircraft's RTA system with the proposed backstepping-based CBF guarantees safety with expected behavior.
First, the safety filter intervenes and makes the aircraft accelerate, pitch up and turn left to avoid collision with the other aircraft.
Then, the aircraft is forced to turn right to avoid crossing the two geofence boundaries.
In this case the safety filter never stops intervening, as the aircraft keeps moving parallel to the geofence rather than returning to the original desired trajectory.
Throughout the motion, the backstepping-based CBF $\hb$ is kept nonnegative, which results in maintaining the three position-based CBFs $\hpi$ (and their smooth under-approximation $\hp$) nonnegative too, as highlighted by Fig.~\ref{fig:sim_backstepping}(b).
These indicate that the underlying maneuvers are executed with guaranteed safety.
While attitude and input constraints are not enforced by the controller, its parameters were tuned so that Euler angles $\theta$, $\phi$ and inputs $P$, $Q$, $\AT$ in Fig.~\ref{fig:sim_backstepping} evolve within reasonable limits.

\subsection{Model-free Run-time Assurance}
\label{sec:modelfree}

The main challenge of designing an RTA system is the synthesis of a CBF that respects the cascaded structure of the underlying dynamics.
This difficulty would vanish if we could use the position-based CBF candidate $\hp$ directly in a model-free fashion.
Next, we discuss a model-free RTA idea, originally introduced in~\cite{molnar2022modelfree} for robotic systems, and describe its benefits and drawbacks relative to the cascaded model-based RTA presented above.

The proposed model-free approach relies on the ability of the aircraft to track a commanded velocity.
Velocity tracking (or trajectory tracking in general) is well-established for many aircraft, for example, the desired controller $\kd$ itself may often be a tracking controller.
Such tracking controllers solve a stabilization problem (associated with Lyapunov theory) that is closely related to ensuring safety (associated with CBF theory)~\cite{molnar2022modelfree}.
As a matter of fact, the tracking controller $\kd$ in Appendix~\ref{appdx:velocity_tracking} is constructed through a backstepping procedure like the safety filter above.
Thus, we intend to leverage the tracking controller (that already accounts for the cascaded dynamics) to address safety in a model-free fashion without going through a complicated control design procedure again.

To ensure safety, we design a safe velocity $\vs(\rr,t)$ to be tracked, by considering the simplified model-independent kinematics ${\dot{\rr} = \vv}$ and regarding $\vv$ as input.
In particular, we synthesize the safe velocity $\vs(\rr,t)$ so that it satisfies:
\begin{equation}
    \dot{h}_{\rm p} \big( \rr, t, \vs(\rr,t) \big) \geq - \alpha_{\rm p} \big( \hp(\rr,t) \big) + \sigma \bigg\| \derp{\hp}{\rr}(\rr,t) \bigg\|^2,
\label{eq:safe_velocity}
\end{equation}
with some ${\sigma > 0}$; cf.~\eqref{eq:safety_condition_velocity}.
The additional term with $\sigma$ originates from the theory of input-to-state safety~\cite{ames2019issf, Alan2022}, and it is incorporated in order to provide robustness against tracking errors when the safe velocity $\vs$ is being tracked.

We use constraint~\eqref{eq:safe_velocity} to synthesize the safe velocity $\vs$ from a desired velocity command $\vvd$ via a smooth safety filter:
\begin{align}
    \vs(\rr,t) & =
    \vvd(\rr,t) + \Lambda \big( a_{\vv}(\rr,t), \| \bb_{\vv}(\rr,t) \| \big) \WW_{\vv} \bb_{\vv}(\rr,t)^\top, \nonumber \\
    a_{\vv}(\rr,t) & = \dot{h}_{\rm p} \big( \rr,t,\vvd(\rr,t) \big) \!+\! \alpha_{\rm p} \big( \hp(\rr,t) \big) \!-\! \sigma \bigg\| \derp{\hp}{\rr}(\rr,t) \bigg\|^2, \nonumber \\
    \bb_{\vv}(\rr,t) & = \derp{\hp}{\rr}(\rr,t) \WW_{\vv},
\label{eq:safe_velocity_smooth}
\end{align}
with $\Lambda$ in~\eqref{eq:safetyfilter_smooth} and a smoothing parameter ${\nu_{\vv} > 0}$; cf.~\eqref{eq:safetyfilter}.
Matrix $\WW_{\vv}$ and ${\GGamma_{\vv} = \WW_{\vv}^{-\top} \WW_{\vv}^{-1}}$ weigh the different velocity components when considering the deviation between $\vs$ and $\vvd$.
For example, if we penalize the deviation of $\vs$ from $\vvd$ in the direction parallel to $\vvd$ and in the perpendicular direction, respectively, with weights $1$ and $\Gamma_{\vv}$, then:
\begin{align}
\begin{split}
    \GGamma_{\vv} & = \PP_{\vv} + \Gamma_{\vv} (\II - \PP_{\vv}), \\
    \WW_{\vv} & = \PP_{\vv} + \frac{1}{\sqrt{\Gamma_{\vv}}} (\II - \PP_{\vv}), \quad
    \PP_{\vv} = \frac{\vvd \vvd^\top}{\| \vvd \|^2}.
\end{split}
\end{align}

Equation~\eqref{eq:safe_velocity_smooth} represents model-free RTA in the sense that it does not use expressions from model~\eqref{eq:Dubins_affine} but only the position-based CBF candidate $\hp$ in~\eqref{eq:CBF_coll}-\eqref{eq:logsumexp}.
As opposed, controllers with the extended CBF~\eqref{eq:extended_CBF_def}-\eqref{eq:extended_CBF} and the backstepping-based CBF~\eqref{eq:Dubins_CBF_backstepping} depended on the model.
Tracking the safe velocity $\vs(\rr,t)$ obtained from the model-free RTA yields safe behavior under certain assumptions about the tracking controller, as established below.
Note that making $\vs$ to be smooth with the smooth safety filter formula facilitates velocity tracking.

\begin{assumption} \label{assum:tracking}
Controller ${\uu = \kk(\xx,t)}$ yields exponentially stable tracking of the safe velocity $\vs(\rr,t)$.
That is, there exist a Lyapunov function ${V\!: \R^n \!\times\! \R_{\geq 0} \!\to \R_{\geq 0}}$ and ${\lambda\!>\!0}$ such that:
\begin{align}
    V(\xx,t) & \geq \frac{1}{2} \| \vs(\rr,t) - \vv(\zzeta) \|^2,
    \label{eq:Lyapunov} \\
    \dot{V} \big( \xx, t, \kk(\xx,t) \big) & \leq - \lambda V(\xx,t),
    \label{eq:Lyapunov_derivative}
\end{align}
where the derivative of $V$ is taken along~\eqref{eq:Dubins_affine}.
\end{assumption}
\noindent
For example, the controller in Appendix~\ref{appdx:velocity_tracking} satisfies Assumption~\ref{assum:tracking}.
Then, it can be established that the system that tracks $\vs$ is safe if $\alpha_{\rm p}$ is chosen such that ${\alpha_{\rm p}(r) = \gamma_{\rm p} r}$ and ${\gamma_{\rm p} < \lambda}$.

\begin{theorem}
\textit{
If the safe velocity $\vs(\rr,t)$ in~\eqref{eq:safe_velocity_smooth} with ${\alpha_{\rm p}(r) = \gamma_{\rm p} r}$ is tracked by~\eqref{eq:Dubins_affine} with a controller ${\uu = \kk(\xx,t)}$ such that~\eqref{eq:Lyapunov}-\eqref{eq:Lyapunov_derivative} hold with ${\gamma_{\rm p} < \lambda}$, then the following set $\S_{V}$ is forward invariant along the closed-loop dynamics~\eqref{eq:closedloop}:
\begin{align}
\begin{split}
    & \S_{V} = \{(\xx,t) \in \R^n \times \R_{\geq 0} : \hV(\xx,t) \geq 0 \}, \\
    & \hV(\xx,t) = \hp(\rr,t) - \frac{V(\xx,t)}{2 \sigma (\lambda - \gamma_{\rm p})}.
\end{split}
\label{eq:CBF_modelfree}
\end{align}
This guarantees that ${(\xx_0,0) \in \S_{V} \implies (\xx(t),t) \in \S_{V},\ \forall t \geq 0}$, which leads to ${(\rr(t),t) \in \S,\ \forall t \geq 0}$.
}
\end{theorem}

\noindent
Note that the restriction ${(\xx_0,0) \in \S_{V}}$ on the initial condition is stricter than ${(\rr_0,0) \in \S}$ due to the term $V(\xx_0,0)/(2 \sigma (\lambda - \gamma_{\rm p}))$ in the expression of $\hV(\xx_0,0)$.
However, the magnitude of this term can be reduced by increasing $\sigma$.

\begin{figure}[!t]
\centering
\includegraphics[scale=1]{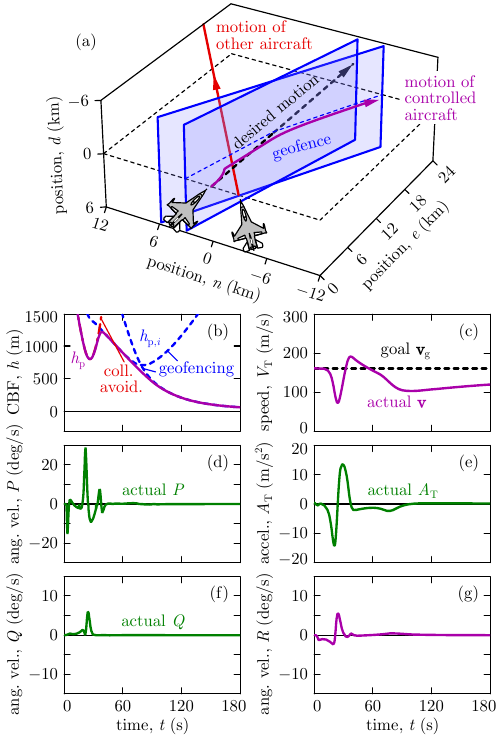}
\caption{Simulation of RTA in simultaneous collision avoidance and geofencing using a model-free approach with position-based CBF.
Safety is maintained, while control inputs are larger than with the model-based method in Fig.~\ref{fig:sim_backstepping}.}
\label{fig:sim_modelfree}
\end{figure}

\begin{proof}
The forward invariance of $\S_{V}$ follows from Lemma~\ref{lemma:CBF}, by showing that the following inequality holds:
\begin{align}
    & \dot{h}_{V} (\xx,t,\kk(\xx,t)) + \gamma_{\rm p} \hV(\xx,t) \nonumber \\
    & = \dot{h}_{\rm p} \big( \rr, t, \vv(\zzeta) \big) \!+\! \gamma_{\rm p} \hp(\rr,t) \!-\! \frac{\dot{V}(\xx,t,\kk(\xx,t))}{2 \sigma (\lambda - \gamma_{\rm p})} \!-\! \frac{\gamma_{\rm p} V(\xx,t)}{2 \sigma (\lambda - \gamma_{\rm p})} \nonumber \\
    & \geq \sigma \bigg\| \derp{\hp}{\rr}(\rr,t) \bigg\|^2 \!\!\!+\! \dot{h}_{\rm p} \big( \rr, t, \vv(\zzeta) \big) \!-\! \dot{h}_{\rm p} \big( \rr, t, \vs(\rr,t) \big) \!+\! \frac{V(\xx,t)}{2 \sigma} \nonumber \\
    & \geq \sigma \bigg\| \derp{\hp}{\rr}(\rr,t) \bigg\|^2 + \derp{\hp}{\rr}(\rr,t) \big( \vv(\zzeta) - \vs(\rr,t) \big) + \frac{V(\xx,t)}{2 \sigma} \nonumber \\
    & \geq \sigma \bigg\| \derp{\hp}{\rr}(\rr,t) \bigg\|^2 - \bigg\| \derp{\hp}{\rr}(\rr,t) \bigg\| \sqrt{2 V(\xx,t)} + \frac{V(\xx,t)}{2 \sigma} \nonumber \\
    & \geq \bigg( \sqrt{\sigma} \bigg\| \derp{\hp}{\rr}(\rr,t) \bigg\| - \sqrt{\frac{V(\xx,t)}{2 \sigma}} \bigg)^2 \nonumber \\
    & \geq 0,
\end{align}
where first we used~\eqref{eq:CBF_modelfree};
second we substituted~\eqref{eq:safe_velocity} and~\eqref{eq:Lyapunov_derivative};
then we expressed the difference of the $\dot{h}_{\rm p}$ terms;
next we applied the Cauchy-Schwartz inequality and~\eqref{eq:Lyapunov};
and finally we completed the square.
Furthermore, safety w.r.t.~$\S_{V}$ implies safety w.r.t.~$\S$, ${(\xx,t) \in \S_{V} \implies (\rr,t) \in \S}$, since ${\hp(\rr,t) \geq \hV(\xx,t)}$ if ${\gamma_{\rm p} < \lambda}$.
\end{proof}

The performance of the model-free RTA system is demonstrated in Fig.~\ref{fig:sim_modelfree}.
The same simultaneous collision avoidance and geofencing task is executed as in Fig.~\ref{fig:sim_backstepping}.
Model~\eqref{eq:Dubins_affine} is simulated with the tracking controller from Appendix~\ref{appdx:velocity_tracking}, which is used to track the safe velocity resulting from the model-free smooth safety filter~\eqref{eq:safe_velocity_smooth}.
The parameters are listed in Table~\ref{tab:parameters}.
The model-free RTA successfully maintains safety in both collision avoidance and geofencing, by keeping the smooth under-approximation $\hp$ of the underlying CBF candidates $\hpi$ nonnegative.
Note that the aircraft with the model-free RTA only uses deceleration and turning to avoid collisions, and it does not move up or down.
This is due to symmetry (and not due to the lack of capability to leverage pitching): the controlled aircraft moves in the same horizontal plane as the other aircraft, since moving up or down would be indifferent because the model-free safety filter does not contain any term like gravity to break the symmetry.

The advantage of this model-free approach is that RTA becomes simpler, and it leverages the existing flight controller (i.e., the desired tracking controller).
Furthermore, this approach does not modify the low-level flight control system that flies the aircraft with desired stable behavior, but only uses high-level commands about which direction and how fast the aircraft should fly.
The disadvantage of the model-free approach is that it is hard to tune it in a way that the safe velocity is easy to track.
This can be observed in Fig.~\ref{fig:sim_modelfree}(d)-(e) which highlight that tracking the safe velocity results in larger control inputs than the ones with the model-based RTA in Fig.~\ref{fig:sim_backstepping}(d)-(e).

In summary, the velocity-based extended CBF in Section~\ref{sec:extended} may succeed in collision avoidance but may fail geofencing tasks, because it is unable to leverage all control inputs (like roll motion) to guarantee safety.
The backstepping-based CBF in Section~\ref{sec:backstepping} and the model-free RTA in Section~\ref{sec:modelfree} both leverage all control inputs and can successfully accomplish simultaneous collision avoidance and geofencing.
The model-based RTA with backstepping  may be preferable over the model-free RTA in terms of the magnitude of control inputs.
Of course, incorporating model information enhances the performance of the control design, but at the price of complexity---this is a fundamental trade-off observed in many control systems.
Nevertheless, the resulting controllers enjoy the formal safety guarantees provided by CBF theory, with both the model-based and model-free RTA.

\subsection{Validation with a Full-Envelope System Identified Model}

\begin{figure}[!t]
\centering
\includegraphics[scale=1]{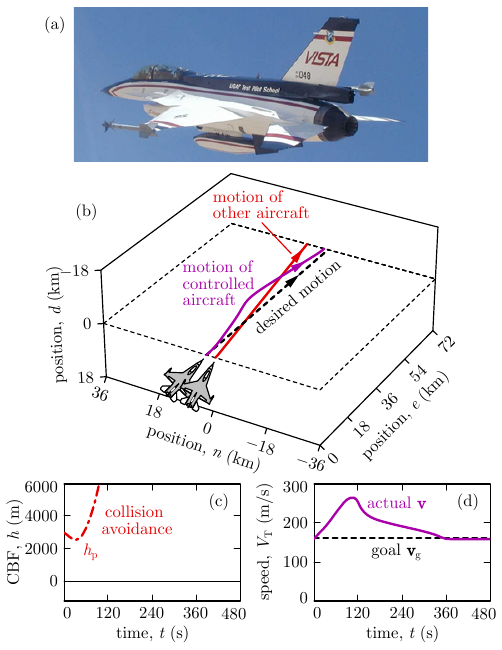}
\caption{Validation simulation of the proposed RTA on a high-fidelity model.
(a) Variable-stability In-flight Simulator Test Aircraft (VISTA), whose linearized, stitched dynamical model from~\cite{Knapp2018} was simulated.
(b,c,d) Simulation of RTA in collision avoidance.
The VISTA tracks the acceleration and angular velocity commands obtained from the RTA that uses 3D Dubins model and backstepping-based CBF.
Safety is maintained in high-fidelity simulation.}
\label{fig:vista}
\end{figure}

Finally, to validate that the RTA developed for model~\eqref{eq:Dubins_affine} could provide safety on a higher-fidelity model, simulations were run using an implementation of the linearized, stitched model of the X-62 VISTA (modified two-seat F-16 aircraft) shown in Fig.~\ref{fig:vista}(a).
This is a proprietary dynamical model that was developed based on \cite{Knapp2018}, where flight test data were used to identify and combine (i.e., stitch together) linear state-space models with time delays that represent the six-degrees-of-freedom dynamics at various flight conditions.
The VISTA simulation was equipped with a neural network controller (NNC) as the nominal flight controller.
The NNC was trained via reinforcement learning~\cite{Hobbs2023, afrl2023dod}, and it provides the necessary commands in terms of power lever angle (PLA), roll rate, and normal load factor to achieve stable wing-level flight at given altitude, heading, and speed. 
With the RTA system proposed in this paper, we seek to modulate the NNC whenever necessary to ensure safety w.r.t.~collision avoidance.
This serves as the first step towards validating that the Dubins model-based RTA system can be executed on more realistic models, while a comprehensive validation, including experiments on hardware, is left for future work.

The high-fidelity simulation results are shown in Fig.~\ref{fig:vista}.
A collision avoidance scenario is considered where the VISTA is initially headed towards another aircraft and an evasive maneuver is required from the RTA.
To track the nominal trajectory, the NNC provides PLA, roll rate, and normal load factor commands which can be converted to desired acceleration and angular velocity (i.e., $\AT$, $P$, and $Q$).
The RTA modifies these desired values to safe acceleration and angular velocity commands.
The safe commands are converted back to PLA, roll rate, and normal load factor, and tracked by the VISTA.
The RTA is based on the 3D Dubins model~\eqref{eq:Dubins_affine} and the backstepping-based CBF~\eqref{eq:Dubins_CBF_backstepping}, which was previously demonstrated in~Fig.~\ref{fig:sim_backstepping}.
The corresponding simulated trajectories are shown by Fig.~\ref{fig:vista}(b)-(d).
Remarkably, the RTA constructed from the lower-fidelity 3D Dubins model is capable of providing the required evasive maneuver, and it successfully guarantees safety even in a high-fidelity simulation---thanks to a carefully designed CBF in~\eqref{eq:Dubins_CBF_backstepping}.

We remark that the high-fidelity simulation environment was also monitoring additional constraints on the aircraft, such as limits for angle of attack, angle of side slip, pilot $g$ command, roll rate, yaw rate, normal load factor, side load factor, speed, control surface deflections, and their rates.
These constraints include some of the states and inputs of the Dubins model, as well as additional quantities that are not present in the model.
Upon violation of these constraints, the simulation terminated.
By tuning the parameters of the proposed controller, we prevented the violation of these constraints while maintaining safety w.r.t.~collision avoidance.

Our future plan is to formally address these state and input constraints by more complex CBF formulations such as the backup set method~\cite{gurriet2020scalable, singletary2022onboard}.
This method uses a backup controller that could maintain safe behavior with input-constrained conservative maneuvers like turning the aircraft away from unsafe regions.
By forward integrating the closed-loop dynamics and computing the corresponding backup trajectory, safety filters can enforce that this trajectory leads to a backup set without violating safety.
This strategy ultimately provides formal guarantees of safety while satisfying input constraints, and its safety filter may integrate a desired flight controller to achieve a higher performance than using a conservative backup controller directly.
Developing the required backup controllers and sets, however, needs significant future research.

\section{Conclusion}
\label{sec:concl}

In this paper, we developed a run-time assurance system for fixed-wing aircraft that intervenes into the operation of existing flight controllers whenever necessary for the safety of the aircraft.
Specifically, we used control barrier functions (CBFs) to establish controllers with formal safety guarantees for collision avoidance, geofencing and the simultaneous execution thereof.
We established and proved the safety guarantees provided by these controllers, and we demonstrated safe operation by numerical simulation of a nonlinear kinematic aircraft model and a high-fidelity dynamical model.
We highlighted that different choices of CBFs---high-order and backstepping-based CBFs that use model information, and simplified position-based CBF candidates in a model-free context---lead to qualitatively different behaviors while guaranteeing safety.
As future work, we plan to enforce safety w.r.t.~other constraints in flight envelopes such as angle of attack bounds or attitude constraints.
We also plan to formally address and enforce input constraints with techniques like the backup set method~\cite{gurriet2020scalable}.
Finally, we plan to validate safety-critical controllers by hardware experiments.



{\appendices
\section{Equations of Motion}
\label{appdx:model}

Here we derive the equations of the 3D Dubins model~\eqref{eq:Dubins}-\eqref{eq:Dubins_R}.
First, we revisit a six-degrees-of-freedom (6 DoF) model from~\cite{stevens2016} that governs the rigid-body dynamics of fixed-wing aircraft.
Then, we simplify these dynamics to the 3D Dubins model through certain assumptions.
While the upcoming equations are well-known in the literature, they are required by the controllers in the main body of the paper.

Consider the aircraft illustrated in Fig.~\ref{fig:model}.
To describe the aircraft's motion, we rely on the
north, east, down position coordinates, $n$, $e$, $d$;
the roll, pitch, yaw Euler angles, $\phi$, $\theta$, $\psi$;
the speed, $\VT$, angle of side slip, $\beta$, angle of attack, $\alpha$;
and the front, right, down angular velocities, $P$, $Q$, $R$:
\begin{equation}
    \rr =
    \begin{bmatrix}
    n \\ e \\ d
    \end{bmatrix}, \quad
    \xxi =
    \begin{bmatrix}
    \phi \\ \theta \\ \psi
    \end{bmatrix}, \quad
    \eeta =
    \begin{bmatrix}
    \VT \\ \beta \\ \alpha
    \end{bmatrix}, \quad
    \oomega =
    \begin{bmatrix}
    P \\ Q \\ R
    \end{bmatrix}.    
\end{equation}
This yields 12 states that evolve according to the 6 DoF model.

To derive the governing equations of motion, we use three reference frames:
the Earth frame, ${\rm e}$, to express global position;
the body frame, ${\rm b}$, aligned with the aircraft's body;
and the wind frame, ${\rm w}$, aligned with the velocity vector of the center of mass.
The orientation of the body frame relative to the Earth frame is given by the Euler angles $\xxi$, whereas the orientation of the wind frame relative to the body frame is described by the angle of side slip $\beta$ and angle of attack $\alpha$.
The frames are related by the transformation matrices:
\begin{align}
\begin{split}
    \Reb(\xxi) & =
    \begin{bmatrix}
        \c_{\psi} \!&\! - \s_{\psi} \!&\! 0 \\
        \s_{\psi} \!&\! \c_{\psi} \!&\! 0 \\
        0 \!&\! 0 \!&\! 1
    \end{bmatrix} \!\!
    \begin{bmatrix}
        \c_{\theta} \!&\! 0 \!&\! \s_{\theta} \\
        0 \!&\! 1 \!&\! 0 \\
        - \s_{\theta} \!&\! 0 \!&\! \c_{\theta}
    \end{bmatrix} \!\!
    \begin{bmatrix}
        1 \!&\! 0 \!&\! 0 \\
        0 \!&\! \c_{\phi} \!&\! - \s_{\phi} \\
        0 \!&\! \s_{\phi} \!&\! \c_{\phi}
    \end{bmatrix}\!, \\
    \Rbw(\eeta) & =
    \begin{bmatrix}
        \c_{\alpha} \!&\! 0 \!&\! - \s_{\alpha} \\
        0 \!&\! 1 \!&\! 0 \\
        \s_{\alpha} \!&\! 0 \!&\! \c_{\alpha}
    \end{bmatrix} \!\!
    \begin{bmatrix}
        \c_{\beta} \!&\! - \s_{\beta} \!&\! 0 \\
        \s_{\beta} \!&\! \c_{\beta} \!&\! 0 \\
        0 \!&\! 0 \!&\! 1
    \end{bmatrix}\!, \\
    \Rbe(\xxi) & = \Reb(\xxi)^\top, \quad
    \Rwb(\eeta) = \Rbw(\eeta)^\top,
\end{split}
\end{align}
where $\c_{(.)}$ abbreviates $\cos(.)$ and $\s_{(.)}$ abbreviates $\sin(.)$.

\subsection{Kinematics}

First, we characterize the kinematics of the aircraft through the evolution of its position and orientation.
The position $\rr$ evolves according to the expression of the velocity $\vv$:
\begin{equation}
    \dot{\rr} = \vv(\xxi,\eeta).
\end{equation}
The velocity of the aircraft's center of mass can be given in the Earth, body, and wind frames, respectively, by:
\begin{align}
\begin{split}
    \vv(\xxi,\eeta) & = \Reb(\xxi) \vb(\eeta), \\
    \vb(\eeta) & = \Rbw(\eeta) \vw(\eeta), \\
    \vw(\eeta) & =
    \begin{bmatrix}
    \VT & 0 & 0
    \end{bmatrix}^\top.
\end{split}
\end{align}

The Euler angles $\xxi$ are related to the angular velocity $\oomega$ by the Euler angle kinematics~\cite{stevens2016}:
\begin{equation}
    \dot{\xxi} = \HH(\xxi) \oomega, \quad
\label{eq:Euler}
\end{equation}
where the coefficient matrix is:
\begin{equation}
    \HH(\xxi) =
    \begin{bmatrix}
    1 & \s_{\phi} \t_{\theta} & \c_{\phi} \t_{\theta} \\
    0 & \c_{\phi} & - \s_{\phi} \\
    0 & \s_{\phi} / \c_{\theta} & \c_{\phi} / \c_{\theta}
    \end{bmatrix},
\label{eq:Euler_matrix}
\end{equation}
and $\t_{\theta}$ abbreviates $\tan(\theta)$.
Moreover, after laborious calculation, it can be shown that the following useful identities hold:
\begin{align}
\begin{split}
    & \derp{\vv}{\xxi}(\xxi,\eeta) \HH(\xxi) \oomega = \Reb(\xxi) \big( \oomega \times \vb(\eeta) \big), \\
    & \derp{\vv}{\eta}(\xxi,\eeta) = \Reb(\xxi) \Rbw(\eeta)
    \begin{bmatrix}
    1 & 0 & 0 \\
    0 & \VT & 0 \\
    0 & 0 & \VT \c_{\beta}
    \end{bmatrix}.
\end{split}
\label{eq:dv_identity}
\end{align}
These expressions define the underlying acceleration:
\begin{equation}
    \dot{\vv} = \derp{\vv}{\xxi}(\xxi,\eeta) \dot{\xxi} + \derp{\vv}{\eta}(\xxi,\eeta) \dot{\eeta},
\label{eq:acceleration}
\end{equation}
where the formula of $\dot{\eeta}$ is introduced below. 

The kinematics simplify as follows in the special case ${\beta \equiv 0}$ and ${\alpha \equiv 0}$.
The body-wind frame transformation reduces to ${\Rwb(\eeta) = \Rbw(\eeta) = \II}$, hence the velocity becomes:
\begin{equation}
    \vv(\xxi,\eeta) = \Reb(\xxi)
    \begin{bmatrix}
        \VT \\ 0 \\ 0
    \end{bmatrix} =
    \begin{bmatrix}
        \VT \c_{\theta} \c_{\psi} \\
        \VT \c_{\theta} \s_{\psi} \\
        -\VT \s_{\theta}
    \end{bmatrix}.
\end{equation}
That is, the velocity only depends on the state $\zzeta$ defined in~\eqref{eq:posvel}, which is emphasized by an abuse of notation in the main body of the paper: $\vv(\zzeta)$ is used instead of $\vv(\xxi,\eeta)$.
Accordingly, ${\derp{\vv}{\phi}(\xxi,\eeta) = \derp{\vv}{\beta}(\xxi,\eeta) = \derp{\vv}{\alpha}(\xxi,\eeta) = \OO}$ holds, and the acceleration in~\eqref{eq:acceleration} reads:
\begin{equation}
    \dot{\vv} = \MM_{\aa}(\xxi,\eeta)
    \begin{bmatrix}
    \AT \\ Q \\ R
    \end{bmatrix},
\label{eq:Dubins_acceleration}
\end{equation}
where ${\AT = \dot{V}_{\rm T}}$ and:
\begin{multline}
    \MM_{\aa}(\xxi,\eeta) =
    \begin{bmatrix}
    \derp{\vv}{\VT}(\xxi,\eeta) \!&\!
    \begin{bmatrix}
    \derp{\vv}{\theta}(\xxi,\eeta) \!&\! \derp{\vv}{\psi}(\xxi,\eeta)
    \end{bmatrix}
    \HH_{\theta \psi}(\xxi)
    \end{bmatrix} \\
    =
    \begin{bmatrix}
    \c_{\theta} \c_{\psi} \!\!&\!\! - \VT (\c_{\phi} \s_{\theta} \c_{\psi} \!+\! \s_{\phi} \s_{\psi}) \!\!&\!\! \VT (\s_{\phi} \s_{\theta} \c_{\psi} \!-\! \c_{\phi} \s_{\psi}) \\
    \c_{\theta} \s_{\psi} \!\!&\!\! \VT (-\c_{\phi} \s_{\theta} \s_{\psi} \!+\! \s_{\phi} \c_{\psi}) \!\!&\!\! \VT (\s_{\phi} \s_{\theta} \s_{\psi} \!+\! \c_{\phi} \c_{\psi}) \\
    -\s_{\theta} \!\!&\!\! - \VT \c_{\phi} \c_{\theta} \!\!&\!\! \VT \s_{\phi} \c_{\theta}
    \end{bmatrix}\!\!,
\label{eq:Dubins_acceleration_M}
\end{multline}
with $\HH_{\theta \psi}$ being the bottom right ${2 \times 2}$ block of $\HH$ in~\eqref{eq:Euler_matrix}.

\subsection{Dynamics}

Next, we describe the dynamics of the aircraft through the evolution of its velocity and angular velocity.
These are related to the forces and moments acting on the body: the thrust force, $\FFT$, assumed to be aligned with the body with magnitude $\FT$;
the aerodynamics forces, $\FA$, including the lift, $L$, drag, $D$, and crosswind, $C$, components in wind frame;
the gravitational force $m \ggD$ with the mass $m$ and the gravitational acceleration $\ggD$ aligned with the down axis in Earth frame with magnitude $\gD$;
and the moments $\MM$ exerted on the aircraft, with body-frame components $L$, $M$, $N$:
\begin{equation}
    \FFT \!=\!
    \begin{bmatrix}
    \FT \\ 0 \\ 0
    \end{bmatrix}\!, \;\;
    \FA(\eeta) \!=\!
    \begin{bmatrix}
    D(\eeta) \\ C(\eeta) \\ L(\eeta)
    \end{bmatrix}\!, \;\;
    \ggD \!=\!
    \begin{bmatrix}
    0 \\ 0 \\ \gD
    \end{bmatrix}\!, \;\;
    \MM \!=\!
    \begin{bmatrix}
    L \\ M \\ N
    \end{bmatrix}\!.
\end{equation}

The acceleration of the aircraft's center of mass can be given in the Earth frame by:
\begin{equation}
    \dot{\vv} = \frac{1}{m} \Reb(\xxi) \big( \FFT - \Rbw(\eeta) \FA(\eeta) \big) + \ggD.
\label{eq:acceleration_vs_force}
\end{equation}
Expressing $\dot{\eeta}$ from~\eqref{eq:acceleration} and substituting~\eqref{eq:Euler}-\eqref{eq:dv_identity} and~\eqref{eq:acceleration_vs_force}
lead to the wind-axes force equations:
\begin{equation}
    \dot{\eeta} = \ff_{\eeta}(\xxi,\eeta,\oomega) + \gg_{\eeta}(\eeta) \FT, \\
\label{eq:force}
\end{equation}
where:
\begin{align}
\begin{split}
    & \ff_{\eeta}(\xxi,\eeta,\oomega) = \AA(\eeta) \Big( \!-\! \frac{1}{m} \FA(\eeta) + \Rwb(\eeta) \Rbe(\xxi) \ggD \\
    & \hspace{30mm} + \vw(\eeta) \times \big( \Rwb(\eeta) \oomega \big) \Big), \\
    & \gg_{\eeta}(\eeta) \!=\! \AA(\eeta) \frac{1}{m} \Rwb(\eeta) \!\!
    \begin{bmatrix}
    1 \\ 0 \\ 0
    \end{bmatrix}\!\!, \quad
    \AA(\eeta) \!=\!
    \begin{bmatrix}
    1 \!\!&\!\! 0 \!\!&\!\! 0 \\
    0 \!\!&\!\! \frac{1}{\VT} \!\!&\!\! 0 \\
    0 \!\!&\!\! 0 \!\!&\!\! \frac{1}{\VT \c_{\beta}}
    \end{bmatrix}\!\!.
\end{split}
\end{align}

Furthermore, the body-frame angular acceleration is:
\begin{equation}
    \dot{\oomega} = \ff_{\oomega}(\oomega) + \gg_{\oomega} \MM, \\
\label{eq:moment}
\end{equation}
where $\ff_{\oomega}$ and $\gg_{\oomega}$ include the mass moment of inertia $\JJ$:
\begin{equation}
    \ff_{\oomega}(\oomega) = - \JJ^{-1} \big( \oomega \times (\JJ \oomega) \big), \quad
    \gg_{\oomega} = \JJ^{-1}.
\end{equation}

\subsection{Governing Equations}

The kinematic and dynamic equations finally lead to the following {\em 6 DoF model}:
\begin{align}
\begin{split}
    \dot{\rr} & = \vv(\xxi,\eeta), \\
    \dot{\xxi} & = \HH(\xxi) \oomega, \\
    \dot{\eeta} & = \ff_{\eeta}(\xxi,\eeta,\oomega) + \gg_{\eeta}(\eeta) \FT, \\
    \dot{\oomega} & = \ff_{\oomega}(\oomega) + \gg_{\oomega} \MM.
\end{split}
\label{eq:6DoF}
\end{align}

We simplify the 6 DoF model to the 3D Dubins model using the following assumptions.
\begin{assumption} \label{assum:model}
The angle of attack, the angle of side slip, and the crosswind force are zero: ${\alpha \equiv 0}$, ${\beta \equiv 0}$, and ${C(\eeta) \equiv 0}$.
The dynamics of the angular velocity $\oomega$ are neglected.
\end{assumption}
\noindent Furthermore, we use the longitudinal acceleration ${\AT = \dot{V}_{\rm T}}$ directly instead of the thrust force $\FT$.
Then, the dynamics of $\beta$, $\alpha$, and $\oomega$ are dropped, leading to the {\em 3D Dubins model}:
\begin{align}
\begin{split}
    \dot{\rr} & = \vv(\xxi,\eeta), \\
    \dot{\xxi} & = \HH(\xxi) \oomega, \\
    \dot{V}_{\rm T} & = \AT,
\end{split}
\end{align}
where ${\eeta = \begin{bmatrix} \VT & 0 & 0 \end{bmatrix}^\top}$.
By spelling out all terms, this gives~\eqref{eq:Dubins}.
Furthermore, since ${\alpha \equiv 0}$, ${\beta \equiv 0}$, and ${C(\eeta) \equiv 0}$, we obtain ${\Rwb(\eeta) = \Rbw(\eeta) = \II}$, and the second component of the body-axes force equations~\eqref{eq:force}, ${\dot{\beta} \equiv 0}$, simplifies to~\eqref{eq:Dubins_R}.
Moreover, the third component of~\eqref{eq:force}, ${\dot{\alpha} \equiv 0}$, gives the required lift force, ${L = m (\gD \c_{\phi} \c_{\theta} + \VT Q)}$, that could keep ${\alpha \equiv 0}$.
This shows that the assumption ${\alpha \equiv 0}$ implies a specific state-dependent lift force.
Although this may not true in practice, we use this assumption as approximation to reduce the full dynamics model to simplified kinematics that represent the aircraft's overall motion with less complexity.

\section{Tracking Controller}
\label{appdx:velocity_tracking}

Finally, we establish a velocity tracking controller for the 3D Dubins model~\eqref{eq:Dubins_affine}.
Note that this controller is independent of the proposed RTA, it is used in simulation examples only, and other tracking controllers
could also be considered.

The velocity tracking controller is designed to track a velocity command $\vc$.
For the model-based RTA examples in Figs.~\ref{fig:sim_extended_coll},~\ref{fig:sim_extended_geo}, and \ref{fig:sim_backstepping}, this command is a desired velocity, ${\vc(\rr,t) = \vvd(\rr,t)}$, chosen to track a goal trajectory $\rr_{\rm g}(t)$:
\begin{equation}
    \vvd(\rr,t) = \vv_{\rm g}(t) + \KK_{\rr} (\rr_{\rm g}(t) - \rr),
\end{equation}
with a symmetric positive definite gain $\KK_{\rr}$, where ${\dot{\rr}_{\rm g} = \vv_{\rm g}}$.
For the model-free RTA example in Fig.~\ref{fig:sim_modelfree}, the safe velocity command ${\vc(\rr,t) = \vs(\rr,t)}$ from~\eqref{eq:safe_velocity_smooth} is tracked.

We construct an exponentially stable tracking controller for the 3D Dubins model~\eqref{eq:Dubins_affine} 
by the help of {\em backstepping} with control Lyapunov functions (CLFs)~\cite{freeman1992backstepping} in two steps:
\begin{enumerate}
    \item We design a desired acceleration $\ad$ that would lead to exponentially stable velocity tracking, with corresponding thrust $\AT$, angular velocity $Q$ (related to pitching), and desired angular velocity $\Rd$ (related to turning).
    $\AT$ and $Q$ are inputs to the 3D Dubins model, making the aircraft accelerate and pitch.
    However, the aircraft cannot be commanded to turn, as $R$ is not an input, and the desired acceleration cannot be realized directly.
    \item We use $\Rd$ in a CLF -- constructed based on backstepping -- to synthesize the remaining input: the angular velocity $P$ (related to rolling).
    Commanding $P$ drives the actual angular velocity $R$ to the desired value $\Rd$, and a desired turning rate is achieved through rolling.
    This ultimately yields exponentially stable velocity tracking.
\end{enumerate}

With this construction, our goal is to establish exponentially stable velocity tracking through the CLF candidate:
\begin{equation}
    V_{0}(\xx,t) = \frac{1}{2} \| \vc(\rr,t) - \vv(\zzeta) \|^2,
\end{equation}
that needs to be driven to zero.
Accordingly, the desired acceleration is designed as:
\begin{equation}
    \ad(\xx,t) = \ac(\xx,t) + \frac{1}{2} \KK_{\vv} \big( \vc(\rr,t) - \vv(\zzeta) \big),
\label{eq:desired_accel}
\end{equation}
where ${\dot{\vv}_{\rm c} = \ac}$ and $\KK_{\vv}$ is a symmetric positive definite gain.
The desired acceleration ensures exponential stability such that the following holds for $\dot{\vv} = \ad(\xx,t)$ and for any ${\lambda > 0}$ that is smaller than or equal to the smallest eigenvalue of $\KK_{\vv}$:
\begin{equation}
    \dot{V}_{0}(\xx,t) = - \frac{1}{2} \big\| \vc(\rr,t) - \vv(\zzeta) \big\|_{\KK_{\vv}}^2 \leq - \lambda V_{0}(\xx,t).
\label{eq:Lyapunov_step}
\end{equation}

We convert the desired acceleration $\ad$ to the inputs $\AT$, $Q$ and the desired angular velocity $\Rd$ based on~\eqref{eq:Dubins_acceleration}:
\begin{equation}
    \begin{bmatrix}
    \AT \\ Q \\ \Rd
    \end{bmatrix} =
    \MM_{\aa}(\xxi,\eeta)^{-1} \ad(\xx,t).
\label{eq:Dubins_accel_conversion}
\end{equation}
The resulting $\AT$ and $Q$ can be directly commanded as inputs, that leads to the actual acceleration:
\begin{equation}
    \dot{\vv} = \ad(\xx,t) + \MM_{R}(\xxi,\eeta) (R - \Rd),
\label{eq:acceleration_actual}
\end{equation}
where $\MM_{R}$ is the last column of $\MM_{\aa}$ in~\eqref{eq:Dubins_acceleration_M}.
Meanwhile, we can use $\Rd$ to design the remaining input $P$ that does not appear in the derivative of $V_{0}$.
This means backstepping: taking a step back from rolling ($P$) to turning ($\Rd$).

In particular, we use backstepping to construct a CLF~\cite{freeman1992backstepping}:
\begin{equation}
    V(\xx,t) = \frac{1}{2} \| \vc(\rr,t) - \vv(\zzeta) \|^2 + \frac{1}{2 \mu} (R - \Rd)^2,
\label{eq:Dubins_CLF_backstepping}
\end{equation}
with a scaling constant ${\mu > 0}$.
Here $R$ and $\Rd$ are given by~\eqref{eq:Dubins_R} and~\eqref{eq:Dubins_accel_conversion}, while their dynamics are of the form:
\begin{align}
\begin{split}
\dot{R} & = f_{R}(\xx,t) + g_{R}(\xx,t) P, \\
\dot{R}_{\rm d} & = f_{\Rd}(\xx,t) + g_{\Rd}(\xx,t) P,
\end{split}
\label{eq:down_angular_acceleration}
\end{align}
where $f_{R}$, $g_{R}$, $f_{\Rd}$, and $g_{\Rd}$ can be obtained -- through lengthy calculation -- by differentiating~\eqref{eq:Dubins_R} and~\eqref{eq:Dubins_accel_conversion}.

Then, we design the remaining input $P$ by enforcing the condition of exponential stability using CLF theory~\cite{Khalil2002}:
\begin{equation}
    \dot{V} \big( \xx, t, \uu \big) + \lambda V(\xx,t) =
    a_{P}(\xx,t) + b_{P}(\xx,t) P \leq 0,
\label{eq:Dubins_backstepping_constraint}
\end{equation}
where, through substituting~\eqref{eq:acceleration_actual}-\eqref{eq:down_angular_acceleration}, the coefficients become:
\begin{align}
\begin{split}
    a_{P}(\xx,t) & =
     - \frac{1}{2} \big\| \vc(\rr,t) - \vv(\zzeta) \big\|_{\KK_{\vv}}^2 \\
    & \quad
    + \big( \vc(\rr,t) - \vv(\zzeta) \big)^\top \MM_{R}(\xxi,\eeta) (\Rd - R) \\
    & \quad
    + \frac{1}{\mu} (\Rd - R) \big( f_{\Rd}(\xx,t) - f_{R}(\xx,t) \big) \\
    & \quad
    + \frac{\lambda}{2} \| \vc(\rr,t) - \vv(\zzeta) \|^2
    + \frac{\lambda}{2 \mu} (\Rd - R)^2, \\
    b_{P}(\xx,t) & =
    \frac{1}{\mu} (\Rd - R) \big( g_{\Rd}(\xx,t) - g_{R}(\xx,t) \big) .
\end{split}
\label{eq:Lyapunov_coefficient}
\end{align}
Here ${\lambda > 0}$ is chosen such that it is smaller than or equal to the eigenvalues of $\KK_{\vv}$ as suggested above.

Thus, $P$ can be designed, for example, by the following CLF quadratic program:
\begin{align}
\begin{split}
    P =
    \underset{\hat{P} \in \R}{\operatorname{argmin}} & \quad
    \hat{P}^2 \\
    \text{s.t.} & \quad
    a_{P}(\xx,t) + b_{P}(\xx,t) \hat{P} \leq 0,
\end{split}
\end{align}
cf.~\eqref{eq:QP}, that has the explicit solution:
\begin{equation}
    P =
    \begin{cases}
    0 & {\rm if}\ b_{P}(\xx,t) = 0, \\
    \frac{\min\{ 0, -a_{P}(\xx,t)\}}{b_{P}(\xx,t)} & {\rm if}\ b_{P}(\xx,t) \neq 0,
    \end{cases}
\end{equation}
cf.~\eqref{eq:safetyfilter}-\eqref{eq:safetyfilter_max}.
Notice that backstepping ensures that this optimization problem is feasible, i.e., there exists $P$ that satisfies~\eqref{eq:Dubins_backstepping_constraint}.
Namely, even when ${R = \Rd}$, which means that ${b_{P}(\xx,t) = 0}$ and $P$ drops from~\eqref{eq:Dubins_backstepping_constraint}, the inequality still holds because ${a_{P}(\xx,t) \leq 0}$ with the above choice of $\lambda$ based on~\eqref{eq:Lyapunov_coefficient} and~\eqref{eq:Lyapunov_step}.
Ultimately, with the backstepping procedure~\eqref{eq:Dubins_CLF_backstepping} and~\eqref{eq:Dubins_backstepping_constraint} provide exponential stability, and also ensure that~\eqref{eq:Lyapunov}-\eqref{eq:Lyapunov_derivative} hold for ${\vc(\rr,t) = \vs(\rr,t)}$.
}

\bibliographystyle{IEEEtran}
\bibliography{2023_tcst}

\vspace{-10mm}
\begin{IEEEbiography}[{\includegraphics[width=1in,height=1.25in,clip,keepaspectratio]{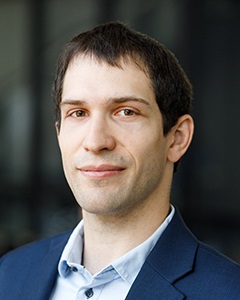}}]
{Tamas G. Molnar} is an Assistant Professor of Mechanical Engineering at the Wichita State University since 2023. Beforehand, he held postdoctoral positions at the California Institute of Technology, from 2020 to 2023, and at the University of Michigan, Ann Arbor, from 2018 to 2020. He received his PhD and MSc in Mechanical Engineering and his BSc in Mechatronics Engineering from the Budapest University of Technology and Economics, Hungary, in 2018, 2015, and 2013. His research interests include nonlinear dynamics and control, safety-critical control, and time delay systems with applications to connected automated vehicles, robotic systems, and autonomous systems.
\end{IEEEbiography}

\vspace{-10mm}
\begin{IEEEbiography}[{\includegraphics[width=1in,height=1.25in,clip,keepaspectratio]{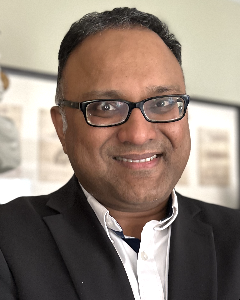}}]
{Suresh K. Kannan} is Chief Scientist at the Nodein Autonomy Corporation. Before Nodein, he led the autonomy group at the United Technologies Research Center (now Raytheon). In 2005, Dr. Kannan received his Ph.D., in Aerospace Engineering from the Georgia Institute of Technology. Between 2000 and 2005 he developed adaptive neural-network-based controllers for DARPA. The real-time adaptation during flight and associated hedging allowed DARPA to safely fly multiple advanced control and planning algorithms. This autonomy stack continues to fly today in academia, commercial drones, and in 2014 on a manned helicopter. Between 2011 and 2017, his work focused on developing vision-based navigation and collision-free trajectories with verifiable guarantees. Between 2014 and 2019, under the DARPA ALIAS program for Sikorsky, he developed a formal language that enables software and human agents to interact verifiably and safely. In other areas, Dr. Kannan helped write the recent 2021 revision of the ASTM F3269 standard for run-time assurance. His singular focus is the practical use of formal mathematics to aid the development of reliable and certifiable algorithms for autonomous aircraft and cars.
\end{IEEEbiography}

\begin{IEEEbiography}[{\includegraphics[width=1in,height=1.25in,clip,keepaspectratio]{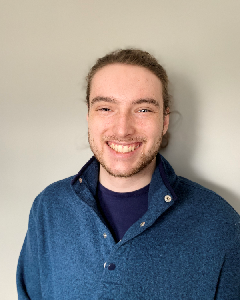}}]{James Cunningham} is an AI Scientist at Parallax Advanced Research, primarily working with the Safe Autonomy Team at the Air Force Research Laboratory's Autonomy Capability Team (ACT3) studying approaches towards safety in autonomous systems which use learning-based control methods. His previous experience includes work in efficient parameter-free online clustering, multi-domain learning-based data embedding and retrieval, automatic supervised large-scale dataset generation, autonomous tracking in Wide Area Motion Imagery, and discriminatory algorithms in the SAR domain. James received his BS and MS degree in Computer Science and Engineering from Ohio State University.
\end{IEEEbiography} 

\begin{IEEEbiography}[{\includegraphics[width=1in,height=1.25in,clip,keepaspectratio]{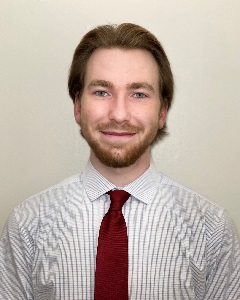}}]{Kyle Dunlap}
is an AI Scientist at Parallax Advanced Research, primarily working with the Safe Autonomy Team at the Air Force Research Laboratory's Autonomy Capability Team (ACT3). There he investigates real time safety assurance methods for intelligent aerospace control systems. His previous experience includes developing a universal framework for Run Time Assurance (RTA) and comparing different RTA approaches during Reinforcement Learning training. He received his BS and MS in Aerospace Engineering from the University of Cincinnati.
\end{IEEEbiography}

\begin{IEEEbiography}[{\includegraphics[width=1in,height=1.25in,clip,keepaspectratio]{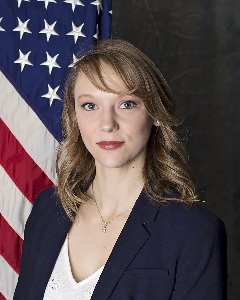}}]{Kerianne L. Hobbs} is the Safe Autonomy and Space Lead on the Autonomy Capability Team (ACT3) at the Air Force Research Laboratory. There she investigates rigorous specification, analysis, bounding, and intervention techniques to enable safe, trusted, ethical, and certifiable autonomous and learning controllers for aircraft and spacecraft applications. Her previous experience includes work in automatic collision avoidance and autonomy verification and validation research. Dr. Hobbs was selected for the 2024 AIAA Associate Fellow Class, 2020 AFCEA 40 Under 40 award, and was a member of the team that won the 2018 Collier Trophy (Automatic Ground Collision Avoidance System Team).  Dr. Hobbs has a BS in Aerospace Engineering from Embry-Riddle Aeronautical University, an MS in Astronautical Engineering from the Air Force Institute of Technology, and a Ph.D. in Aerospace Engineering from the Georgia Institute of Technology. 
\end{IEEEbiography}

\begin{IEEEbiography}[{\includegraphics[width=1in,height=1.25in,clip,keepaspectratio]{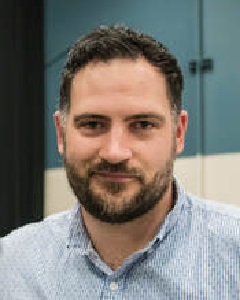}}]
{Aaron D. Ames} is the Bren Professor of Mechanical and Civil Engineering and Control and Dynamical Systems at Caltech. Prior to joining Caltech in 2017, he was an Associate Professor at Georgia Tech in the Woodruff School of Mechanical Engineering and the School of Electrical {\&} Computer Engineering. He received a B.S. in Mechanical Engineering and a B.A. in Mathematics from the University of St. Thomas in 2001, and he received a M.A. in Mathematics and a Ph.D. in Electrical Engineering and Computer Sciences from UC Berkeley in 2006. He served as a Postdoctoral Scholar in Control and Dynamical Systems at Caltech from 2006 to 2008, and began his faculty career at Texas A{\&}M University in 2008. At UC Berkeley, he was the recipient of the 2005 Leon O. Chua Award for achievement in nonlinear science and the 2006 Bernard Friedman Memorial Prize in Applied Mathematics, and he received the NSF CAREER award in 2010, the 2015 Donald P. Eckman Award, and the 2019 IEEE CSS Antonio Ruberti Young Researcher Prize.  His research interests span the areas of robotics, nonlinear, safety-critical control and hybrid systems, with a special focus on applications to dynamic robots -— both formally and through experimental validation.
\end{IEEEbiography}


\end{document}